\documentclass[twocolumn, twocolappendix]{aastex631}
\usepackage{amsmath}

\begin{document}

\title{On the meaning of the dynamo radius in giant planets with stable layers}

\correspondingauthor{Paula N. Wulff}
\email{paulawulff@epss.ucla.edu}

\author[0000-0002-6788-8898]{Paula N. Wulff}
\affiliation{Department of Earth, Planetary, and Space Sciences,\\ 
University of California, Los Angeles (UCLA)\\
595 Charles E Young Dr East, Los Angeles, CA 90095, USA}

\author[0000-0002-6917-8363]{Hao Cao}
\affiliation{Department of Earth, Planetary, and Space Sciences,\\ 
University of California, Los Angeles (UCLA)\\
595 Charles E Young Dr East, Los Angeles, CA 90095, USA}

\author[0000-0002-8642-2962]{Jonathan M. Aurnou}
\affiliation{Department of Earth, Planetary, and Space Sciences,\\ 
University of California, Los Angeles (UCLA)\\
595 Charles E Young Dr East, Los Angeles, CA 90095, USA}

\begin{abstract} 

Current structure models of Jupiter and Saturn suggest that helium becomes immiscible in hydrogen in the outer part of the planets’ electrically conducting regions. This likely leads to a layer in which overturning convection is inhibited due to a stabilizing compositional gradient. The presence of such a stably stratified layer impacts the location and mechanism of convectively-driven dynamo action.

Juno’s measurements of Jupiter’s magnetic field enabled an estimate of its dynamo radius based on the magnetic Lowes spectrum. A depth of $\sim0.8~R_J$ is obtained, where $1~R_J$ is Jupiter’s radius. This is rather deep, considering that the electrical conductivity inside Jupiter is expected to reach significant values at $\sim0.9~R_J$.

Here we use 3-dimensional numerical dynamo simulations to explore the effects of the existence and location of a stably stratified helium rain layer on both the inferred Lowes radius and location of the radial extent of dynamo action. We focus on a Jupiter-like internal structure and electrical conductivity profile. We find that for shallower stable layers, there is no magnetic field generation occurring above the stable layer and the effective dynamo radius and the inferred Lowes radius is at the base of the layer. For deeper stable layers, Lowes radii of $\sim0.87~R_J$ are inferred as a shallow secondary dynamo operates above the stable layer. Our results strongly suggest the existence of a stable layer extending from $\sim0.8~R_J$ up to at least $\sim0.9~R_J$ inside Jupiter. The physical origin of this extended stable layer and its connection to helium rain remain to be elucidated.

\end{abstract}


\section{Introduction} \label{sec:intro}

Our solar system hosts six planets with active dynamos: the terrestrial planets Earth and Mercury; the gas giants Jupiter and Saturn; and the ice giants Uranus and Neptune \citep[e.g., see a recent review article by][]{Soderlund_2025}. When comparing the measured magnetic fields at their surfaces, the differences are not only due to the characteristics of the respective planetary dynamos, but also the distance to the source region. The radial location of the outmost extent of the dynamo region is commonly referred to as the dynamo radius. A method introduced by \citet{Lowes_1974}, and applied to the Earth, enables the inference of the outmost radial extent of dynamo action in a given planet, using the surface and/or space measurements of its magnetic field. The approach relies on two assumptions: 1) that the magnetic field spectrum at the top of the dynamo region is flat; and 2) that the layers above the dynamo region, extending to the surface, are current-free so that the magnetic field there can be described via a scalar potential. If these assumptions are fulfilled, the inferred Lowes radius should give a good approximation of the planet's dynamo radius.

However, the application of this method to gas giants and the meaning of the depths derived are more delicate than for Earth. In the Earth there is a clear boundary between the mantle and the outer core. In contrast, a gas planet's electrical conductivity increases smoothly with depth, across the transition from molecular to metallic hydrogen in the case of the gas giants and water/ammonia/methane ionization in the ice giants. Further, the likely presence of stably stratified layers near the dynamo radius adds another source of complexity. In Jupiter and Saturn there likely exists a region where helium is immiscible in hydrogen \citep{Stevenson_1977, Brygoo_2021}, which is expected to lead to inhibition of large-scale overturning convection. In the ice giants there may be a similar layer where water is immiscible \citep{Amoros_2024}, although \citet{Bailey_2021} argue on thermodynamic grounds that hydrogen and water are separated so that immiscibility would lead to a phase change boundary between the two.

\begin{figure*}[t]
\centering
\includegraphics[width=1.0\linewidth]{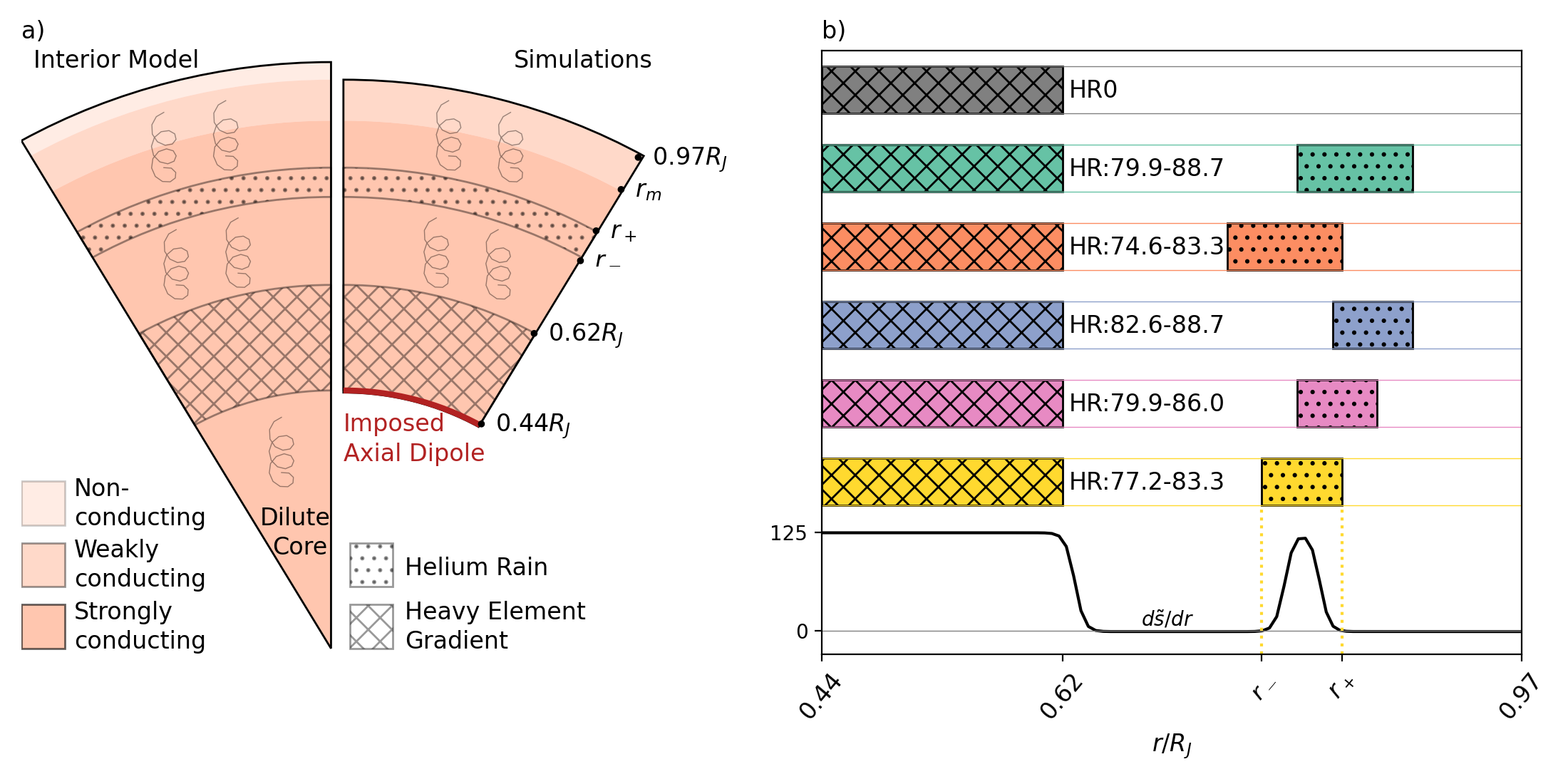}
\caption{a) schematic showing interior structure of Jupiter based on \citet{Militzer_2024} (5-layer reference model) in the left slice, and our numerical dynamo model set-up in the right slice. b) the areas of stable stratification in the various models, with the model names (tab.~\ref{tab:Results}) - referring to the helium rain layer extent - labeled. \label{fig:Int_Sim}}
\end{figure*}

The Juno mission has delivered measurements of Jupiter's magnetic field with incredible spatial resolution \citep{Connerney_2018, Moore_2018, Connerney_2022, Bloxham_2022}. The Juno reference model through perijove 33 \citep[JRM33,][]{Connerney_2022} has resolution up to spherical harmonic degree 18, which allows for an estimation of the Lowes radius inside Jupiter, which has also been loosely referred to as the jovian dynamo radius by some authors. An application of the Lowes method to Jupiter led to estimates of $0.807\pm0.006\,R_J$ by \citet{Connerney_2022} and $0.830\pm0.022\,R_J$ by \citet{Sharan_2022}, where 1 $R_J$ is Jupiter's 1-bar equatorial radius, $71,492~km$. These different estimates rely on different parts of the spectrum as well as on different magnetic field models. Here, we focus our numerical dynamo modeling with a Jupiter-like interior and make comparisons to the JRM33 magnetic field model.


Revised models for Jupiter's interior structure, using Juno's gravity measurements, feature multiple layers as opposed to the classical idea of a fully convective sphere with a central rocky core. The idea of helium rain - a region where helium becomes immiscible in hydrogen - leading to a layer of stable stratification due to compositional gradient, had been proposed for Saturn for many years \citep{Stevenson_1977}. Evidence for the same phenomenon occurring in Jupiter, albeit to a lesser extent, was brought by the measurement of a depletion of helium (and neon) by the Galileo probe \citep{vonZahn1998,WilsonMilitzer2010}, as well as by experimental evidence by \citet{Brygoo_2021}. In addition to the helium rain layer, the central region of the planet may feature a gradient in heavy elements, as proposed by \citet{Stevenson_1985} and recently being further refined using Juno measurements \citep{Wahl_2017, Debras_2019, Miguel_2022, Militzer_2022, Howard_2023, Militzer_2024}. This could lead to a central dilute core, hosting convection and possibly dynamo generation, with a constant mass fraction of heavy elements \citep{Militzer_2022,Militzer_2024}. Above this may lie a stably stratified transition region, where the heavy element abundance transitions to the (lower) surface value. This layer would separate an inner dynamo region from a second dynamo region, atop the transition zone (see fig.~\ref{fig:Int_Sim}a).

The exact location, and even existence, of a stable layer due to helium rain or a deep one due to a heavy elements gradient (HEG), is difficult to constrain as illustrated in \citet{Militzer_2024}, due to the degeneracy of the problem allowing for different models with a similar quality of fit to the gravity data. \citet{Militzer_2024} prefer a thick stable layer, between $70-87\%\,R_J$, while \citet{Markham_2024} promote a thinner layer, with a lower limit of only 20 km ($<0.03\%\,R_J$).

The presence of stably stratified layers in Jupiter can significantly impact the magnetic field generation within the planet, as regions which inhibit convection are unlikely to host dynamo action. First, a deep, thick, stable layer separates the convective dynamo region into two layers with an SSL in-between. Second, the two convective dynamo layers can interact with each other across the SSL. Finally, the ``shallower" stable helium rain layer can further impact the morphology of the surface magnetic field. If there are no significant zonal flows present in the helium rain layer, the magnetic field will experience a passive magnetic skin effect, where rapidly time-varying magnetic field fluctuations are attenuated in the layer \citep[e.g.,][]{Christensen_2006, Christensen2018,Gastine_2020}. If there are strong zonal flows in the stable layer, the magnetic field will be strongly axisymmetrised, as non-axisymmetric components will be filtered out more effectively in the presence of zonal flows \citep{Stevenson_1982, Christensen_2008}.

There are several existing studies attempting to model the Jovian dynamo using 3D magnetohydrodynamic (MHD) simulations. Several of them include a radially varying electrical conductivity, representing the transition from the non-conducting, molecular hydrogen envelope, to the fully conducting metallic hydrogen interior \citep{Heimpel_2011, Duarte_2013, Gastine_2014, Duarte_2018}. However, there are few studies incorporating stable layers. The first to include a stably stratified helium rain layer in the context of Jupiter was \citet{Gastine_2021}. The layer was implemented between $0.84-0.88\,R_J$, motivated by \citet{Militzer_2016} and \citet{Wahl_2017}. However, the resultant surface magnetic field was too axisymmetric compared to that of Jupiter. The authors suggested a shallower helium rain layer would yield a more Jupiter-like field.

\citet{Moore_2022} included a stably stratified dilute core \citep[in contrast to the fully convective dilute core proposed by][]{Militzer_2022} as well as a helium rain layer in 3D numerical dynamo models for Jupiter. The helium rain layer in their model was located between $0.8-0.95\,R_J$, with thicknesses of $0.05, 0.1, 0.15\,R_J$. The surface spectra of these models were compared with the \citet{Connerney_2018} jovian power spectrum. However, the inferred depth of the dynamo region was not calculated. Their favored models have a helium rain layer between $0.9-0.95\,R_J$, which is shallower compared to that adopted in \citet{Gastine_2021}.

Both \citet{Gastine_2021} and \citet{Moore_2022} also sought to reproduce Jupiter's alternating surface zonal wind profile by the inclusion of a helium rain layer. Analysis of the latest Jupiter gravity field models point to a wind depth $\sim2,000-2,500~$km if the deep wind features similar amplitude to the surface wind \citep{Kaspi2023,Cao2023ApJ}. The theoretical work by \citet{Christensen_2024} illustrates that a stable layer responsible for quenching the zonal winds should be located at the inferred depth of deep wind, $\sim2,000-2,500~$km, which corresponds to $\sim$ 0.97 $R_J$ and is much shallower than the depths at which helium rain is expected. This implies that there could be a very wide stable region, with a top boundary of $\sim0.97\,R_J$ inferred by the zonal wind truncation analysis, directly connected to the deeper helium rain stable stratification. Alternatively, there could be two separate layers with convection occurring in between. 

\citet{Tsang_2020} conducted a numerical dynamo study applying the Lowes method for inferring the dynamo radius (the location of the top of the dynamo region) for a fully-convective Jupiter. They found that, in a fully convective jovian interior, the top of the dynamo region is directly linked to the electrical conductivity profile. They obtained Lowes radii of $0.86-0.88\,R_J$. They suggested that the discrepancy between their results and the radii inferred using the Jupiter spectra is likely to be linked to the presence of a stable layer in the outer envelope, similar to the physics considered in this study. However, they did not conduct any numerical dynamo models with a stably stratified layer. 

Here, we build on the extensive work exploring the subtleties of this method to infer the dynamo radius of the Earth by the geodynamo community \citep[e.g.,][]{Lowes_1974, Langel_1982, Cain_1989, Langlais_2014} and apply it to our numerical dynamo models of Jupiter. We explore how the location of a stable layer, relative to the radially varying electrical conductivity, influences the inferred value of the Lowes radius. Furthermore, we evaluate if the inferred Lowes radius agrees with the location of the top of the dynamo of the respective model. We discuss the application of the method to giant planet magnetic field measurements and show that it may also be applied to multipolar magnetic fields.

\section{Dynamo radius and Lowes radius}\label{sec:LowesRad}

Here we first discuss a physically motivated definition of dynamo radius and then outline the Lowes method to infer a planetary dynamo radius (the Lowes radius) from magnetic field measurements.

\begin{figure}[t]
\centering
\includegraphics[width=1.0\linewidth]{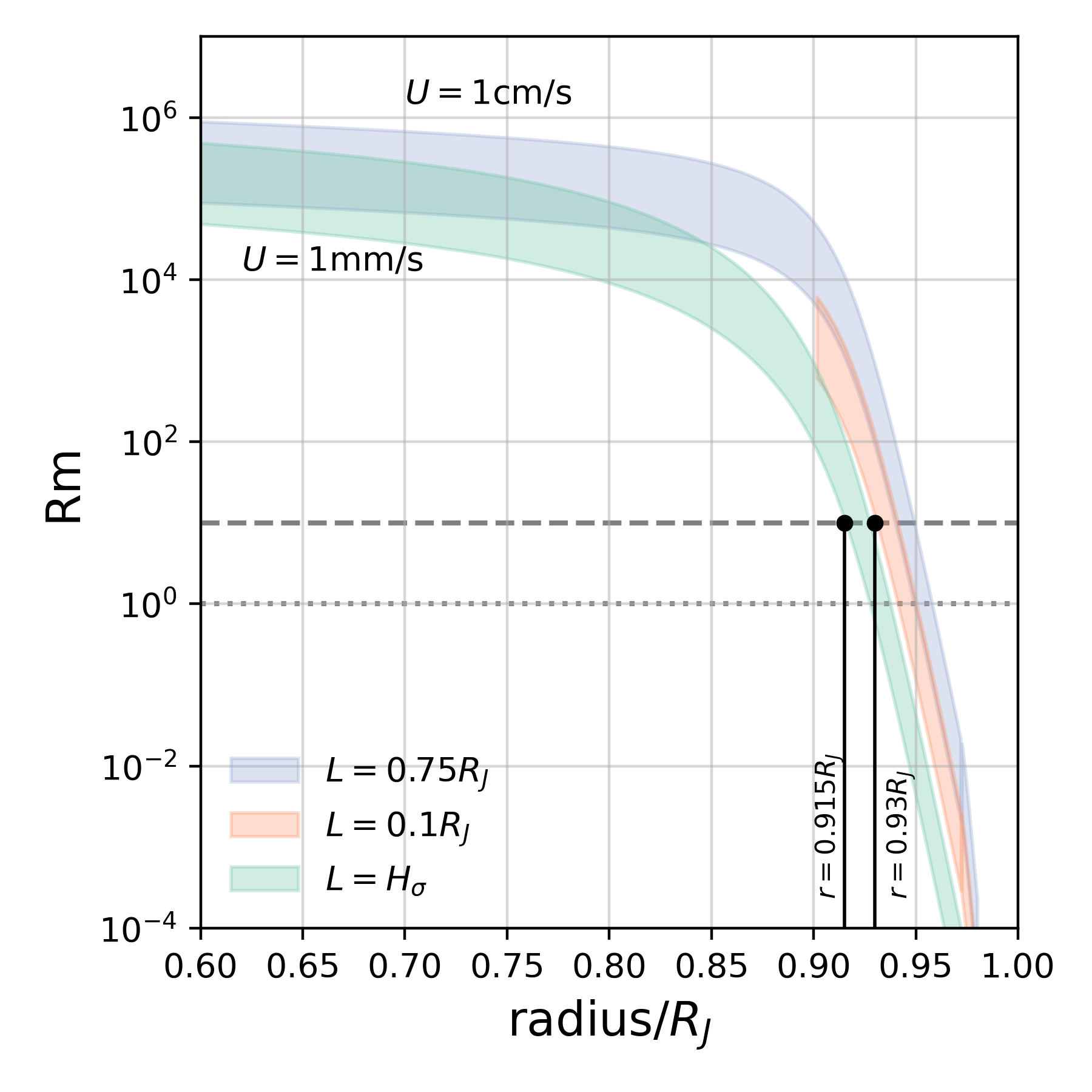}
\caption{Estimation of local magnetic Reynolds number as a function of radius for Jupiter. The lower and upper bound of each envelope is based on a velocity of $1~mm/s$ and $1~cm/s$, respectively. The colors represent three different length-scales used in the definition of $Rm$. Two conservative estimates of the dynamo radius, based on where $Rm$ reaches 10 corresponding to $1~mm/s$ velocity, $L=0.1~R_J$ and $L=H_\sigma$, are marked with horizontal dashed line and dotted line. \label{fig:RmJup}}
\end{figure}


\subsection*{The dynamo radius in a planetary interior}\label{sec:Rm}

The ratio of magnetic induction to diffusion, evaluated locally or globally, is the typical measure for potential dynamo action. In a given system the ratio is approximated by the magnetic Reynolds number, which is generally defined as $Rm=UL/\lambda$, where $U$ and $L$ are characteristic velocities and length-scales, respectively, and $\lambda=1/\mu\sigma$ is the magnetic diffusivity of the fluid, where $\sigma$ is the electrical conductivity and $\mu$ is the magnetic permeability of the fluid. In the deep dynamo region, a bulk $Rm\gtrsim50$ is required to achieve self-sustained dynamo action, based on the findings of numerical dynamo surveys \citep{Kutzer_2002, Christensen_2006b, Gastine_2014}. When $Rm$ is less than unity, only a small modification of the background field is expected.

For giant planets in which electrical conductivity can vary strongly with depth, the choice of the characteristic length-scale, $L$, is non-trivial. \citet{Liu_2008} showed that, with regards to induction due to zonal flows, $\overline{u}_\phi$, the electrical conductivity scale height, $H_\sigma=|\sigma(d\sigma/dr)^{-1}|$, is the smallest length-scale in the outer weakly conducting region, yielding the definition:
\begin{equation}
    Rm_\phi = u_\phi H_\sigma/\lambda.
\end{equation}
However, when considering the radial component of the magnetic induction equation \citep[e.g.][]{Amit_2007},
\begin{equation}
    \frac{\partial B_r}{\partial t}=-\boldsymbol{u}_h\cdot\nabla B_r-B_r\nabla_h\cdot\boldsymbol{u}_h+\frac{1}{\mu_0\sigma}\hat{r}\cdot\nabla^2 \boldsymbol{B},
\end{equation}
where subscript $h$ denotes the horizontal direction, we observe that the radial derivative of the electrical conductivity does not feature in the ratio of induction to diffusion of radial magnetic field. Therefore, when evaluating the magnetic Reynolds number associated with $B_r$, the thickness of the dynamo region $D$ may be a more appropriate length-scale, and the characteristic radial velocity a more appropriate velocity scale. Another intermediate length-scale would be $\sqrt{H_\sigma D}$.

Fig.~\ref{fig:RmJup} shows the radial profiles of a range of possible appropriate magnetic Reynolds numbers for Jupiter. We use the electrical conductivity profile fitted to the ab initio calculations of \citet{French_2012}. An upper estimate of $1~cm/s$ for velocity is adopted based on zonal flow speeds from \citet{Bloxham_2022} and a lower estimate for convective velocities of $1~mm/s$ \citep[see Table~1 in][]{Starchenko_2002}, indicated by the top and bottom bounds of each envelope. The colors indicate three different length-scales used: $L=0.75~R_J$ (thickness of a large, fully convective dynamo region), $L=0.1~R_J$ (thickness of an upper dynamo region, bounded by a stable layer below), and $H_\sigma(r)$ (electrical conductivity scale-height).

Taking the conservative value of $1~mm/s$ for the flow speed, the local magnetic Reynolds number is expected to reach 10 at $\sim91.5\%~R_J$ based on $L=H_\sigma$ or $\sim93\%~R_J$ based on $L=0.1~R_J$, respectively. Thus, we expect dynamo action inside Jupiter to reach a substantial level at a depth of $\sim$0.92 $R_J$ if there are no ``shallow" stable layers present. 


\subsection*{Inferring the dynamo radius via the Lowes method}

The Lowes method is the main approach to infer the dynamo radius of a planet from surface and space magnetic field measurements. We refer the reader to \citet{Mauersberger_1956, Lowes_1966, Lowes_1974} for derivations and introductions to the concept. \citet{Tsang_2020} apply the method to (fully convective) numerical dynamo simulations with radially varying electrical conductivity. \citet{Connerney_2022, Sharan_2022} conduct a heuristic application to measurements at Jupiter and \citet{Langlais_2014} apply it to all outer solar system planets.

\begin{figure}[t]
\centering
\includegraphics[width=1.0\linewidth]{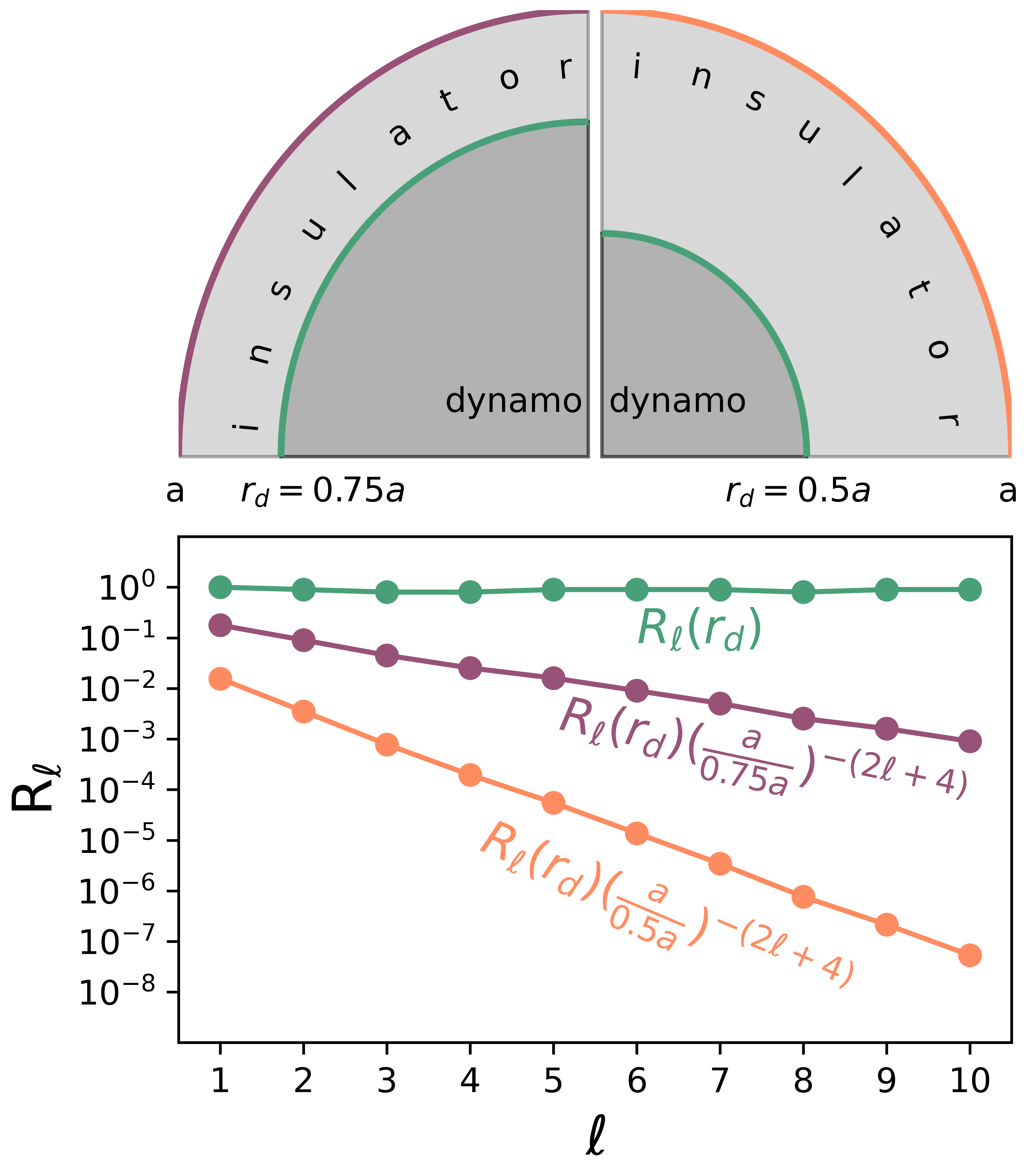}
\caption{Schematic illustrating the effect of passing through a non-electrically conducting zone on a magnetic spectrum. This idealised schematic assumes that the \textit{inferred} dynamo radius $r_{lowes}$ is equal to the actual dynamo radius $r_d$. \label{fig:LowesSchem}}
\end{figure}

The basic approach relies on the assumption that in a turbulent planetary dynamo, magnetic field/energy is generated at all spatial scales equally, referred to as the ``White Source Hypothesis" \citep{Backus_1996}. It follows that this corresponds to a flat, or white, spectrum throughout the dynamo region. This is equivalent to assuming that the magnetic energy is randomly distributed across different length-scales. Commonly, this randomness is attributed to turbulence in the dynamo region. However, there exists no quantitative connection between rotating MHD turbulence in the planetary dynamo region and the assumed white spectrum. Although this argument is heuristic, we will show with our 3D numerical dynamo models that it holds reasonably well in the dynamo regions (at least in our models). Such a flat spectrum at the top of the dynamo region is shown schematically in Fig.~\ref{fig:LowesSchem} (green).

If the planetary dynamo is hidden from the surface by a non-conducting outer envelope (the mantle in the case of the Earth; molecular hydrogen in the case of gas giants), this upper region will be current-free ($\mathbf{j}=0$) and the field at the top of the dynamo region will follow the radial dependence of a scalar potential, $\Psi$:
\begin{equation}
    \nabla\times\mathbf{B}=\mu_0\mathbf{j}=0\,\rightarrow\,\boldsymbol{B}=-\nabla\Psi,
\end{equation}
all the way to the surface. Since $\nabla\cdot\mathbf{B}=0$, it follows that
\begin{equation}
    \nabla^2\Psi=0.
\end{equation}
\citet{Gauss_1839} showed that, for a magnetic source located below the spherical surface at which measurements are made, the potential may be written as:
\begin{align}
    \Psi(r,\theta,\phi)=a\sum^\infty_{\ell=1}&\sum^\ell_{m=0}\left(\frac{a}{r}\right)^{\ell+1}P_{\ell}^m(\cos\theta)\nonumber\\
    &\cdot[g_\ell^m\cos(m\phi)+h_\ell^m\sin(m\phi)],
\end{align}
where $r,\theta,\phi$ are the radial, colatitudinal and longitudinal coordinates and $a$ is a reference radius. Either the mean radius or the equatorial radius of the planet is usually adopted for $a$. $P_{\ell}^m$ are the Schmidt semi-normalized associated Legendre polynomials while $g_{\ell m}$ and $h_{\ell m}$ are the Gauss coefficients, measured at the planetary surface (defined by the reference radius). $\ell$ and $m$ represent spherical harmonic degree and order, respectively.

\citet{Mauersberger_1956} and \citet{Lowes_1966} defined the spatial power spectrum of the geomagnetic field (applicable to any planet). This is commonly referred to as the Lowes-Mauersberger spectrum, or simply the Lowes spectrum, and represents the magnetic energy for every spherical harmonic degree $\ell$ integrated over the spherical surface $\mathbf{\Omega}$ at radius $r$:
\begin{align}
    R_\ell(r)=& \int_\mathbf{\Omega} |\mathbf{B}_\ell (r)|^2 d\mathbf{\Omega}=\langle |\boldsymbol{B}_\ell (r)|^2\rangle \nonumber\\
    =&(\ell+1)\left(\frac{a}{r}\right)^{2\ell+4}\sum^\ell_{m=0}[(g_\ell^m)^2+(h_\ell^m)^2]\nonumber\\
    =& R_\ell(a)\left(\frac{a}{r}\right)^{2\ell+4},\label{eq:Lowes}
\end{align}
where $\langle\cdot\rangle$ denotes a surface integral. Thus, we can write the squared magnetic field strength, integrated over a radial level in this region as a sum of these terms:
\begin{equation}
    \langle |\boldsymbol{B}(r)|^2\rangle=\sum^{\ell_{max}}_{\ell=1}R_\ell(a)\left(\frac{a}{r}\right)^{2\ell+4}.
\end{equation}

Invoking the White Source Hypothesis then yields the following expression, linking the field at the \textit{inferred} top of the dynamo region, $r_{Lowes}$, with that at the surface:
\begin{equation}
    R_\ell(r_{Lowes})=\text{const.}= R_\ell(a)\left(\frac{a}{r_{Lowes}}\right)^{2\ell+4}.
\end{equation}
The schematic shown in Fig.~\ref{fig:LowesSchem} illustrates what the magnetic spectrum at the surface of a planet, shown in purple (orange), may look like if the Lowes radius is located at 75\% (50\%) of the planetary radius.

We take the base 10 logarithm of eq.~(\ref{eq:Lowes}) and differentiate with respect to degree $\ell$:
\begin{equation}
    \frac{d\log_{10}R_\ell(r)}{d\ell}=\frac{d\log_{10}R_\ell(a)}{d\ell}+2\log_{10}\frac{a}{r},\label{eq:specderiv}
\end{equation}
and use $\beta$ to express the amplitude of the \textit{expected} local slope of the spectrum in the current-free region:
\begin{equation}
    \beta(r>r_{Lowes})\sim-\frac{d\log_{10}R_\ell(r>r_{Lowes})}{d\ell}.
\end{equation}
Any base for the logarithm may be used here and we follow \citet{Tsang_2020} with the choice of $\log_{10}$. For the measurable spectrum at the surface this gives:
\begin{equation}
    \log_{10} R_\ell(a)\sim-\beta(a)\ell.\label{eq:betasurf}
\end{equation}
Thus, we can substitute these expressions into eq.~(\ref{eq:specderiv}) and write the expected slope of the spectrum as a function of depth, $\beta(r)$, within $r_{Lowes}<r<a$:
\begin{equation}
    \beta(r)=\beta(a)-2\log_{10}\frac{a}{r}.\label{eq:betar}
\end{equation}

Solving eq.~(\ref{eq:betar}) for $\beta(r_{Lowes})=0$ then yields the inferred Lowes dynamo radius, $r_{Lowes}$:
\begin{equation}
    r_{Lowes}=10^{-\beta(a)/2}a,\label{eq:rlowes}
\end{equation}
Thus, this method allows one to use external $B_r$ measurements to infer the outmost radial extent of the dynamo region, as long as the two main assumptions are satisfied. 

While for the planets we only have measurements at or above their surfaces, our numerical dynamo models contain information at all radial levels. Consequently, the local slope of the magnetic energy spectrum can be evaluated as a function of radius:
\begin{equation}
    \alpha(r)\sim-\frac{d\log_{10} R_\ell(r)}{d\ell},
\end{equation}
and an equivalent expression of eq.~\ref{eq:betasurf} at any radial level is:
\begin{equation}
    \log_{10} R_\ell(r)\sim-\alpha(r)\ell. \label{eq:alpha}
\end{equation}
We can evaluate $\alpha(r)$ for all models in this study, and test how much it deviates from the expected slope, $\beta(r)$, above the Lowes radius under the current-free assumption.

The range of spherical harmonic degrees used to make the fitting with the surface spectrum varies between different studies in the geodynamo community. Typically, low degrees are excluded, in particular the dipole $\ell=1$, as they deviate from the flat spectrum at the core-mantle boundary. This is also the case for the Jupiter analysis as \citet{Connerney_2022} omit $\ell=1,2$ and make the fit over $\ell=3-18$. \citet{Langlais_2014} find that two families of spectral contributions fit the White Source Hypothesis at the Earth: the non-axisymmetric components, and the quadrupole family modes where $\ell+m=\text{an even number}$. We will compare the results of the Lowes method analysis using both total and non-axisymmetric contributions, with a varying range of $\ell$'s.

\section{Methods}

We construct and analyze a suite of 3D numerical dynamo studies to investigate the bulk dynamics of dynamo action inside Jupiter-like planets with stable layers (e.g., see Fig. ~\ref{fig:Int_Sim}). 

\subsection{MHD Equations}\label{sec:eqns}

The MHD equations are solved under the anelastic approximation \citep{Jones_2011, Gastine_2012}, governing the conservation of mass 
\begin{equation}
    0 = \nabla\cdot(\tilde{\rho}\boldsymbol{u}),
\end{equation}
and momentum
\begin{align}
    \frac{\partial\boldsymbol{u}}{\partial t}&+\boldsymbol{u}\cdot\nabla\boldsymbol{u}+\frac{2}{E}\boldsymbol{e}_z\times\boldsymbol{u}
    = -\nabla\frac{p}{\tilde{\rho}} \nonumber \\
    &+ \frac{1}{E Pm\tilde{\rho}}(\nabla\times\boldsymbol{B})\times\boldsymbol{B} 
     - \frac{Ra}{Pr}s^\prime\tilde{T}\tilde{\alpha}\boldsymbol{g} + \frac{1}{\tilde{\rho}}\nabla\cdot \mathcal{S},
\end{align}
where $\mathcal{S}$ is the rate-of-strain tensor
\begin{equation}
    \mathcal{S}_{ij} = 2\rho\left(\text{e}_{ij}-\frac{1}{3}\frac{\partial u_i}{\partial x_j}\right), \quad \text{e}_{ij}=\frac{1}{2}\left(\frac{\partial u_i}{\partial x_j}+\frac{\partial u_j}{\partial x_i}\right).
\end{equation}
We also solve for the conservation of internal energy
\begin{align}
    \tilde{\rho}\tilde{T}\Big(\frac{\partial s^\prime}{\partial t}+\boldsymbol{u}\cdot\nabla s^\prime& + u_r\frac{d\tilde{s}}{dr}\Big) = \frac{1}{Pr}\nabla\cdot(\tilde{\rho}\tilde{T}\nabla s^\prime) \\
    &+ \frac{Pr Di}{Ra E Pm^2}((\nabla\times\boldsymbol{B})^2 + \frac{Pr Di}{Ra}Q_\nu,
\end{align}
where $Q_\nu$ is viscous heating:
\begin{equation}
    Q_\nu=2\rho\left[\sum_{ij}\text{e}_{ij}\text{e}_{ji}-\frac{1}{3}(\nabla\cdot\boldsymbol{u})^2\right].
\end{equation}
The magnetic induction equation is solved:
\begin{equation}
    \frac{\partial\boldsymbol{B}}{\partial t} = \nabla\times\Big(\boldsymbol{u}\times\boldsymbol{B}-\frac{\tilde{\lambda}}{Pm}\nabla\times\boldsymbol{B}\Big),
\end{equation}
for a radially varying magnetic diffusivity, $\lambda$. Here, $\boldsymbol{u}$ and $\boldsymbol{B}$ are the velocity and magnetic fields, respectively, and $s^\prime$ is entropy fluctuations. Tilde denotes the reference states of density $\rho$, temperature $T$, entropy gradient $ds/dr$, and thermal expansion $\alpha$. The non-dimensionalisation is done using the viscous diffusion time $d^2/\nu$ for time, $\nu/d$ for velocity and $\sqrt{\Omega\mu_0\lambda_i\rho_o}$ for the magnetic field, where subscript $o$ ($i)$ denotes values at the outer (inner) boundary.

The hydrostatic, adiabatic background reference state is defined by the temperature profile
\begin{equation}
    \frac{1}{\tilde{T}}\frac{d\tilde{T}}{dr}=-Di\,\tilde{\alpha}\,\tilde{g},
\end{equation}
A polytropic equation of state is assumed, $\tilde{\rho}(r)=\tilde{T}^n$, where $n$ is the polytropic index. We set the same density contrast across our modelled shell as in \citet{Militzer_2024} (5-layer reference model) where $\tilde{\rho}(0.44R_J)/\tilde{\rho}(0.97R_J)=21$. We choose a polytropic index of $n=2$, with $Di=1.6$, where the dissipation number is defined as
\begin{equation}
    Di=\frac{\alpha_o g_o d}{c_p}.
\end{equation}
This choice also fits well with the background temperature profile in \citet{Militzer_2024}.

The equations are solved with stress-free mechanical boundary conditions and the magnetic field matches a potential at both boundaries. Following \citet{Gastine_2021}, we fix entropy at $r_o$ and fix entropy flux at $r_i$.

\subsection{Hydrodynamic Control Parameters}

We choose a geometric aspect ratio of $\chi=r_i/r_o=0.45$, where $r_i$ represents $\sim0.44R_J$ and $r_o$ represents $0.97R_J$. We cut off our models at $0.97\,R_J$, i.e. the lower extent of the rapid zonal winds observed on the surface, as they do not have a significant impact on the magnetic field. Including the outer few percent where the rapid zonal winds prevail and ensuring that the fluid dynamics there are in a regime dominated by rotation would require extremely high spatial and temporal resolution simulations. The lower extent of our numerical dynamo models represents $0.4365R_J$ as directly modelling the dynamics of the convective dilute core is not relevant to the Lowes radius analysis.

The gravity profile adopted in our model is $g(r)=1/r^2$ and is non-dimensionalised with respect to $g_o$. The dimensionless control parameters are the Prandtl number $Pr$, magnetic Prandtl number $Pm$, Ekman number $E$, and Rayleigh number $Ra$. They are defined as follows:
\begin{equation}
    Pr=\frac{\nu}{\kappa},\, Pm=\frac{\nu}{\lambda_i},\, E=\frac{\nu}{\Omega d^2},\, Ra=\frac{\alpha_o T_o g_o d^4}{c_P \nu\kappa}\left|\frac{d\tilde{s}}{dr}\right|_o,
\end{equation}
where shell thickness is $d=r_o-r_i$. Kinematic viscosity $\nu$ and thermal diffusivity $\kappa$ are both held constant.

For all simulations we fix $Pr=0.5$, $Pm=0.5$, $E=3\times10^{-5}$ and $Ra=1.2\times10^8$. Due to the length-scale dependence of $E$, the effective Ekman number for the convective regions ($E_{eff}$) is larger than the nominal one. The extremes are $E_{eff}=8.5\times10^{-4}$ for the convective region above the helium rain layer, for models with $r_+=0.887R_J$ (see Fig.~\ref{fig:Int_Sim}), and $E_{eff}=4.4\times10^{-4}$ for the convective region between the heavy element gradient region and helium rain layer, for models with $r_-=0.746R_J$. This means that in these particular models the thinner convective regions are less rotationally dominated than we would prefer. However, we believe that $E_{eff}$ does not significantly impact our conclusions concerning the extent of the dynamo region.

\subsection{Stably Stratified Regions}\label{sec:SSLs}

Stably stratified regions are imposed in the models by prescribing an analytical background entropy gradient. In stably stratified regions this has a positive value, which we refer to as $A_{HEG}$ for the stable heavy element gradient (HEG) region and $A_{HR}$ for the stable helium rain (HR) layer. In the convective regions it is fixed at $d\tilde{s}/dr=-1$. The transitions between the regions are smooth and of fixed thickness $\delta=1/75$, for all stable layer boundaries in our models. We implement the HEG stable layer in all models and also add options for an HR layer with a variable extent (see Fig~\ref{fig:Int_Sim}b):
\begin{align}
    \frac{d\tilde{s}}{dr}=& -1+(A_{HEG}+1)\left[1-\frac{1}{2}\left(1-\tanh\frac{r_{HEG} - r}{\delta}\right) \right] \nonumber\\
    &  + 0.25(A_{HR}+1)\left(1+\tanh\frac{r-r_{HR}}{\delta}\right)\nonumber\\
    &\cdot\left(1- \tanh\frac{r -r_{HR} - d_{HR}}{\delta}\right). \label{eq:SSL}
\end{align}
The boundary of the deep stably stratified layer (SSL) lies at $r_{HEG}=1.16\,(=0.62R_J)$, as motivated by \cite{Militzer_2024}. The sandwich layer, representing helium rain, is located between lower boundary $r_{HR}$, and upper boundary $r_{HR}+d_{HR}$. The range of values used can be found in Table~\ref{tab:Results} and is visualized in the schematic in Fig.~\ref{fig:Int_Sim}b. The actual extent of the stably stratified zone is where $d\tilde{s}/dr>0$, thus, the effective extent of each stable region is $\sim0.03(=0.017R_J)$ below the prescribed value of $r_{HR}$ and above the prescribed value of  $r_{HR}+d_{HR}$. We label the upper and lower radial level where $d\tilde{s}/dr$ crosses zero in the helium rain layer $r_+$ and $r_-$, respectively, as shown for the example profile of $d\tilde{s}/dr$ in Fig.~\ref{fig:Int_Sim}b. The set-up allows for different values of $A$ in the upper and lower stable regions. However, we did not judge it a critical parameter to explore for the purposes of this study and therefore set it to $A_{HEG}=A_{HR}=125$ for all models.

The degree of stability of the layers can be evaluated by considering the ratio of the Brunt-V\"ais\"al\"a frequency $N$ (the buoyancy frequency) to rotation rate, $\Omega$, in these regions. The ratio can be found using:
\begin{equation}
    \frac{N}{\Omega} = \sqrt{\tilde{\alpha}(r)\tilde{T}(r)\tilde{g}(r)\frac{Ra E^2}{Pr}\frac{d\tilde{s}}{dr}}.
\end{equation}
The reference $N/\Omega$ used is the average between the respective stable layer boundaries and is roughly equal to 10 in the heavy element gradient region and around 6 in the helium rain layer.

The thickness of the helium rain layer is either $\sim8.8\%$ or $\sim6.1\%$ of Jupiter's radius in our models. The extremely thin layer thickness of 20~km found as a minimum in \citet{Markham_2024} is not reached as this is currently too costly to resolve.

\subsection{Magnetic Parameters}

In our non-dimensionalization framework, the magnetic diffusivity (visualized in Fig.~\ref{fig:Rm}b) is equal to the inverse of the electrical conductivity: $\lambda =1/\sigma$. For the electrical conductivity we use the formula from \citet{GomezPerez_2010}, which describes a near constant conductivity in the metallic hydrogen region ($r<r_m$, see fig.~\ref{fig:Int_Sim}a), dropping off exponentially in the outer molecular hydrogen region.
\begin{equation}\label{eq:Sigma}
    \sigma(r)=\begin{cases}
        1+(\sigma_m-1)\left(\frac{r-r_i}{r_m-r_i}\right)^\zeta, & \text{if $r<r_m$},\\
        \sigma_m\exp\left(\zeta\frac{\sigma_m-1}{\sigma_m}\frac{r-r_m}{r_m-r_i} \right), & \text{if $r_m<r$}.
  \end{cases}
\end{equation}
Adopting the same values as \cite{Gastine_2021}, who fit the electrical conductivity profile from \citet{French_2012}, $\sigma_m=0.07$, $\zeta=11$ and the drop-off radius is adjusted for our re-scaled shell $r_m=0.9R_J=0.93r_o$. To avoid numerical issues we cap $\sigma$ to a minimum value of $=1\times10^{-7}$.

The models of \citet{Militzer_2024} feature an inner dynamo region, hosted inside the convecting dilute core (see schematic in Fig.~\ref{fig:Int_Sim}a). Rather than model this region directly, we cut our models off at the boundary between the convective dilute core and the heavy element gradient region and impose an axial dipole at the inner boundary, motivated by two main considerations. First, implementing a boundary condition as a proxy for an inner dynamo region, rather than including it in the simulation, is computationally cost-saving. Second, there is a significant magnetic skin-effect across the thick stably stratified heavy element gradient region. Therefore, an axial dipole is a good approximation of the primary components of an internal magnetic field generated in the convective dilute core that would reach the overlying convective region.

\subsection{Numerical Methods}

We use version 6.2 of the open-access community dynamo code MagIC, available at \url{https://magic-sph.github.io/}, which makes use of the SHTns library by \citet{Schaeffer_2013}. The anelastic approximation used is described in detail in \citet{Gastine_2012}.

The code solves the equations outlined in sec.~\ref{sec:eqns} where the magnetic $\boldsymbol{B}$ and momentum $\rho\boldsymbol{u}$ fields are expanded into poloidal and toroidal potentials. Chebychev polynomials are used in the radial direction and spherical harmonics in the azimuthal and latitudinal directions. We use $N_r=145$ radial gridpoints and  $N_\phi=1024$ azimuthal gridpoints.

For further details on the implementation of stably stratified layers in MagIC we refer the reader to \citet{Dietrich_2018, Gastine_2021}.

\section{Results}

\begin{figure*}[t!]
\centering
\includegraphics[width=1.0\linewidth]{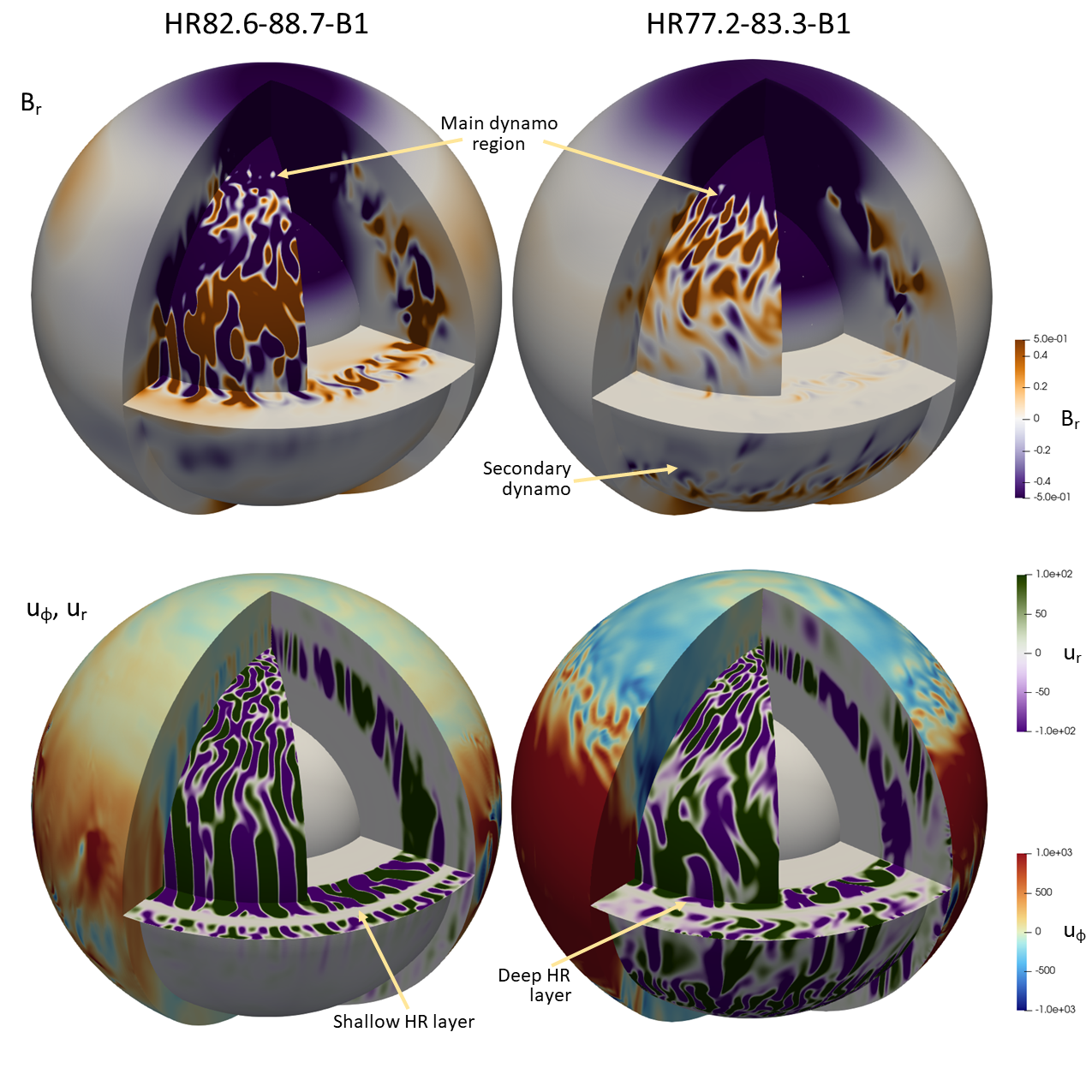}
\caption{3D visualizations of two dynamo models, both with an imposed axial dipole of amplitude 1 with a shallow helium rain layer ($0.826-0.887~R_J$, left) and a deep helium rain layer ($0.772-0.833~R_J$, right). Top panels: radial magnetic field. Bottom panels: radial (green/purple colormap) and zonal (red/blue colormap) velocity. \label{fig:Br_2D}}
\end{figure*}
Fig.~\ref{fig:Br_2D} shows snapshots of models HR82.6-88.7-B1 (left) and HR77.2-83.3-B1 (right). These models have a helium rain with lower boundaries $r_-=0.826~R_J$ and $r_-=0.772~R_J$, and upper boundaries $r_+=0.887~R_J$ and $r_+=0.833~R_J$, respectively, and imposed axial dipole of amplitude 1, (see Table~\ref{tab:Results}). The top panels show their radial magnetic fields while the lower panels show their radial and azimuthal flows. The magnetic field is small-scale and complex below both models' helium rain layers (the figure shows $0.73~R_J$), where there is strong convection. In model HR82.6-88.7-B1 the field is then smoothed out in the shallow stable layer which transitions almost directly into the non-conducting region. Meanwhile, in model HR77.2-83.3-B1, additional magnetic field generation takes place above the helium rain layer as there is space for another region of magnetoconvection.
\begin{table*}[t]
    \centering
    \begin{tabular}{l||ccc||cc||ccc}
         &  & inputs &  & \multicolumn{2}{c}{$r_{Lowes}\,(R_J)$} & & \\
        model & $r_{HR}$ & $d_{HR}$ & $B_{imp}$ & ($\ell=3-18$) & ($\ell=15-35$) & $f_{dip}$ & axisym. & Dipole tilt ($^\circ$) \\
        \hline\hline
        HR0-B0 & -- & -- & 0 & $0.8761\pm0.0075$ & $0.8795\pm0.0010$ & 0.378 & 0.663 & 7.68 \\
        HR0-B0.5 & -- & -- & 0.5 & $0.8716\pm0.0053$ & $0.8761\pm0.0005$ & 0.244 & 0.427 & 14.16 \\
        HR0-B1 & -- & -- & 1 & $0.8434\pm0.0083$ & $0.8769\pm0.0008$ & 0.263 & 0.376 & 18.24 \\
        HR0-B2 & -- & -- & 2 & $0.8588\pm0.0074$ & $0.8809\pm0.0017$ & 0.566 & 0.830 & 4.54 \\
        \hline
        HR79.9-88.7-B0 & 1.53 & 0.1 & 0 & $0.7717\pm0.0138$ & $0.8085\pm0.0013$ & 0.266 & 0.447 & 7.78 \\
        HR79.9-88.7-B0.5 & 1.53 & 0.1 & 0.5 & $0.8040\pm0.0127$ & $0.8054\pm0.0016$ & 0.497 & 0.816 & 12.09 \\
        HR79.9-88.7-B1 & 1.53 & 0.1 & 1 & $0.8156\pm0.0124$ & $0.8179\pm0.0028$ & 0.561 & 0.963 & 3.60 \\
        HR79.9-88.7-B2 & 1.53 & 0.1 & 2 & $0.7985\pm0.0092$ & $0.8094\pm0.004$ & 0.682 & 0.993 & 0.866 \\
        \hline
        HR74.6-83.3-B0 & 1.43 & 0.1 & 0 & $0.9124\pm0.0042$ & $0.9015\pm0.0055$ & 0.640 & 0.975 & 1.17 \\
        HR74.6-83.3-B0.5 & 1.43 & 0.1 & 0.5 & $0.8581\pm0.0105$ & $0.8884\pm0.0062$ & 0.688 & 0.995 & 2.56 \\
        HR74.6-83.3-B1 & 1.43 & 0.1 & 1 & $0.9224\pm0.0051$ & $0.8812\pm0.0051$ & 0.566 & 0.999 & 0.56 \\
        HR74.6-83.3-B2 & 1.43 & 0.1 & 2 & $0.8617\pm0.0037$ & $0.8538\pm0.0108$ & 0.642 & 0.999 & 0.45 \\
        \hline
        HR82.6-88.7-B0 & 1.58 & 0.05 & 0 & $0.8154\pm0.0099$ & $0.8250\pm0.0011$ & 0.036 & 0.297 & 39.08 \\
        HR82.6-88.7-B0.5 & 1.58 & 0.05 & 0.5 & $0.8407\pm0.0043$ & $0.8246\pm0.0008$ & 0.160 & 0.708 & 31.31 \\
        HR82.6-88.7-B1 & 1.58 & 0.05 & 1 & $0.8214\pm0.0082$ & $0.8255\pm0.0007$ & 0.456 & 0.834 & 8.21 \\
        HR82.6-88.7-B2 & 1.58 & 0.05 & 2 & $0.8291\pm0.0093$ & $0.8331\pm0.0034$ & 0.658 & 0.990 & 2.33 \\
        \hline
        HR79.9-86.0-B0 & 1.53 & 0.05 & 0 & $0.8086\pm0.0083$  & $0.8529\pm0.0010$  & 0.119 & 0.408 & 27.67 \\
        HR79.9-86.0-B0.5 & 1.53 & 0.05 & 0.5 & $0.7832\pm0.0079$  & $0.8433\pm0.0032$  & 0.354 & 0.594 & 26.59 \\
        HR79.9-86.0-B1 & 1.53 & 0.05 & 1 & $0.8126\pm0.0070$  & $0.8255\pm0.0019$  & 0.561 & 0.974 & 2.36 \\
        HR79.9-86.0-B2 & 1.53 & 0.05 & 2 & $0.7926\pm0.0054$  & $0.8204\pm0.0027$  & 0.681 & 0.947 & 4.87 \\
        \hline
        HR77.2-83.3-B0 & 1.48 & 0.05 & 0 & $0.9244\pm0.0027$ & $0.8954\pm0.0047$ & 0.333 & 0.974 & 3.72 \\
        HR77.2-83.3-B0.5 & 1.48 & 0.05 & 0.5 & $0.8892\pm0.0102$ & $0.8589\pm0.0075$ & 0.427 & 0.976 & 7.92 \\
        HR77.2-83.3-B1 & 1.48 & 0.05 & 1 & $0.8671\pm0.0107$ & $0.8693\pm0.0041$ & 0.564 & 0.989 & 2.03 \\
        HR77.2-83.3-B2 & 1.48 & 0.05 & 2 & $0.8792\pm0.0021$ & $0.8780\pm0.0022$ & 0.668 & 0.997 & 0.96 \\
        \hline\hline
        Jupiter (JRM33) & -- & -- & -- & $0.8069\pm0.0059$ & -- & 0.740 & 0.770 & 10.25
    \end{tabular}
    \caption{Summary of time-averaged results. The extent of the sub-adiabatic region, $r_--r_+$, is indicated in the first part of each model name (in percentage of $R_J$) while the second part refers to the amplitude of the imposed axial dipole. Values of $r_{HR}$ and $d_{HR}$ (see eq.~\ref{eq:SSL}) are given in the shell coordinate system. $r_{Lowes}$ is given as obtained by fitting to the non-axisymmetric components of the spectrum between $\ell=3-18$ and to the total spectrum between $\ell=15-35$. We also give axial dipolarity, axisymmetry and dipole tilt projected onto $R_J$.}
    \label{tab:Results}
\end{table*}

Table~\ref{tab:Results} summarizes the results of the Lowes analysis for the suite of models as well as three of the key characteristics of their magnetic fields, extrapolated to $1~R_J$. We choose to evaluate the characteristics of the magnetic field at this level, rather than the top of each respective dynamo region (which would be more similar to fully conducting geodynamo studies), as this is then consistent across all our models and can be directly compared to the observed Jovian value at $1~R_J$. \citet{Soderlund_2025b} highlights the variety of definitions used to characterize magnetic field properties in the literature. As our study is not Earth-focused, we choose these three more general definitions, based on the total magnetic power and not only up to $\ell=12$, to represent some key aspects of each model's magnetic field morphology. The dipolarity, $f_{dip}$, is the ratio of magnetic power in the axial dipole component, relative to the total magnetic power. The axisymmetry is the power in the $m=0$ part of the spectrum, relative to the total, and the dipole tilt is also given in degrees.

Almost all of these models can be considered dipolar as $f_{dip}>0.1$ for all stable layer locations and imposed dipole amplitudes, apart from HR82.6-88.7-B0. However, despite imposing an axial dipole field at the lower boundary for most of the models, the value of $f_{dip}$ does not reach the Jupiter value of $f_{dip}=0.740$ at $1~R_J$. This could partly be due to the value of $f_{dip}$ being relative to the magnetic spectrum up to degree $\ell=18$ for Jupiter, while for the simulations it is relative to the full resolution of the model. However, as can be seen qualitatively in Fig.~\ref{fig:Br_2D}, it is also due to the dipolar field deep in the main dynamo region being funneled towards the poles when passing through the helium rain layer. This results in an octupole component which is almost as strong as the dipole component at the surface.

There is a broad range of values for the axisymmetry of the magnetic fields of models without a helium rain layer and with shallow, thin helium rain layers (HR0 and HR82.6-88.7). Models with deep helium rain layers tend to have a high degree of axisymmetry, in particular those with stable layers with an upper boundary of $0.833~R_J$. The azimuthal velocity fields shown in Fig.~\ref{fig:Br_2D} suggest that this is due to the presence of stronger zonal flows in these models, leading to a more efficient filtering out of non-axisymmetric magnetic field components. Furthermore, for the shallow layers the upper part of the stable region lies where electrical conductivity is beginning to decrease, also leading to a diminished EM filtering effect. While it could be expected that dynamo action taking place above the helium rain layer in the deep layer models may re-introduce significant non-axisymmetric components, we observe that these seem to remain so small scale that they are then significantly reduced geometrically (via $1/r^{\ell+2}$) when passing through the non-conducting region and reaching the surface.

While these complex models are rich in physics, we now focus our analysis on the influence of a stably stratified layer on both the inferred and actual dynamo radii of our simulations.

\subsection{Magnetic Energy Spectra}

\begin{figure}[t]
\centering
\includegraphics[width=1.0\linewidth]{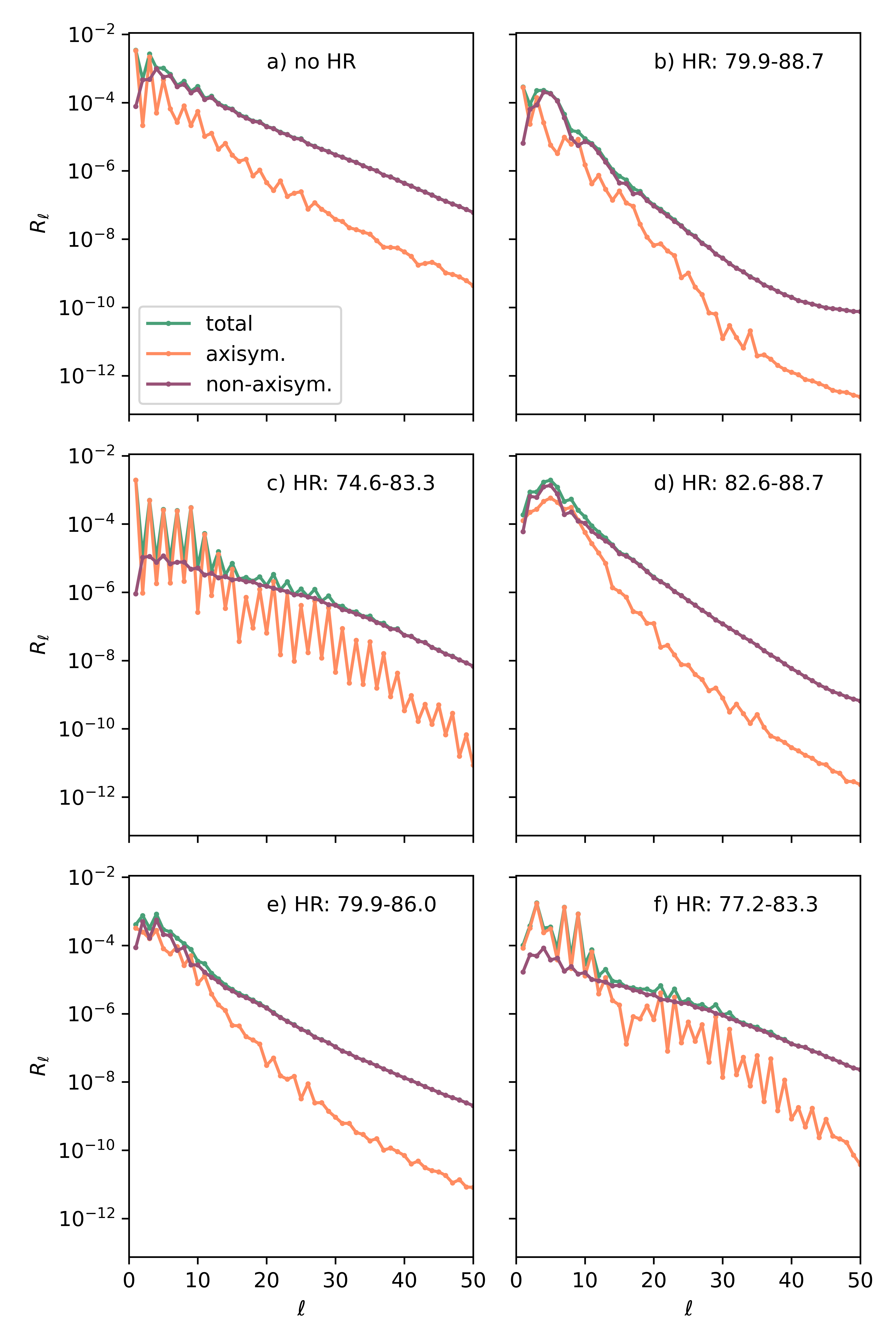}
\caption{Time-averaged Lowes spectra at model surface ($r_o=0.97R_J$) for 6 models with no imposed axial dipole field and varying helium rain layer locations. Green lines correspond to the total spectrum, orange includes only the axisymmetric component, $m=0$, and purple is the non-axisymmetric spectrum. \label{fig:B0_SpecDecomp}}
\end{figure}

The Lowes-Mauersberger spectra of the simulations with no imposed axial dipole are shown in Fig.~\ref{fig:B0_SpecDecomp} with the location of their helium rain layers labeled. We use the time-averaged Lowes spectra for our analysis (not the Lowes spectra of the time-averaged magnetic field), described in more detail in Appendix \ref{sec:AppA}. The total spectrum is shown in green, the axisymmetric ($m=0$) spectrum in orange, and the non-axisymmetric spectrum in purple. All of the spectra are rather irregular at the low degrees. It can be seen that this is mainly due to the axisymmetric component of the field, in particular for the models with deep helium rain layers which are extremely axisymmetric (e.g., c and f in the figure). The non-axisymmetric Lowes spectra are much smoother in comparison, in agreement with analysis of the geomagnetic field \citep{Langlais_2014}. The non-axisymmetric field has no strong overall preference of even $\ell$ versus odd $\ell$, as all length-scales in the non-zonal components get excited and the symmetry properties of non-axisymmetric field are determined by the specific combination of $\ell$ and $m$, hence there are no pronounced zig-zags in the non-zonal spectra. At the higher degrees the axisymmetric component becomes around 2 orders of magnitude smaller than the non-axisymmetric part and the total spectrum (dominated by the non-axisymmetric part) is an almost perfectly straight line on the logarithmic plot.


The shape of the surface spectra motivates us to make the fitting to find the spectral slope (see Eq.~\ref{eq:betasurf}) in two different ways:
\begin{enumerate}
    \item using $\ell=3-18$ of the non-axisymmetric spectrum, as this allows for direct comparison with JRM33. The magnetic fields of our models are generally more axisymmetric than that of Jupiter, thus, unlike \citet{Connerney_2022}, we do not include the $m=0$ component of the spectra.
    \item using $\ell=15-35$, to make use of having a much greater spatial resolution than the Juno measurements and evaluate the effect of including higher degrees when using the Lowes method.
\end{enumerate}

Comparing the spectra shown in Fig.~\ref{fig:B0_SpecDecomp}b and d with the spectrum shown in Fig.~\ref{fig:B0_SpecDecomp}a, it is clear that the spectra in Fig.~\ref{fig:B0_SpecDecomp}b and d are much steeper, indicating that the dynamo source region lies deeper down in the models with a shallow stably stratified layer than in the model with no helium rain layer.

Comparing the spectra shown in Fig.~\ref{fig:B0_SpecDecomp}c and f, with the spectrum shown in Fig.~\ref{fig:B0_SpecDecomp}a, the surface spectra have very similar slopes and indeed, the fitting leads to a very similar value for $r_{Lowes}$. 
\subsection{Lowes radii}

\begin{figure}[t]
\centering
\includegraphics[width=1.0\linewidth]{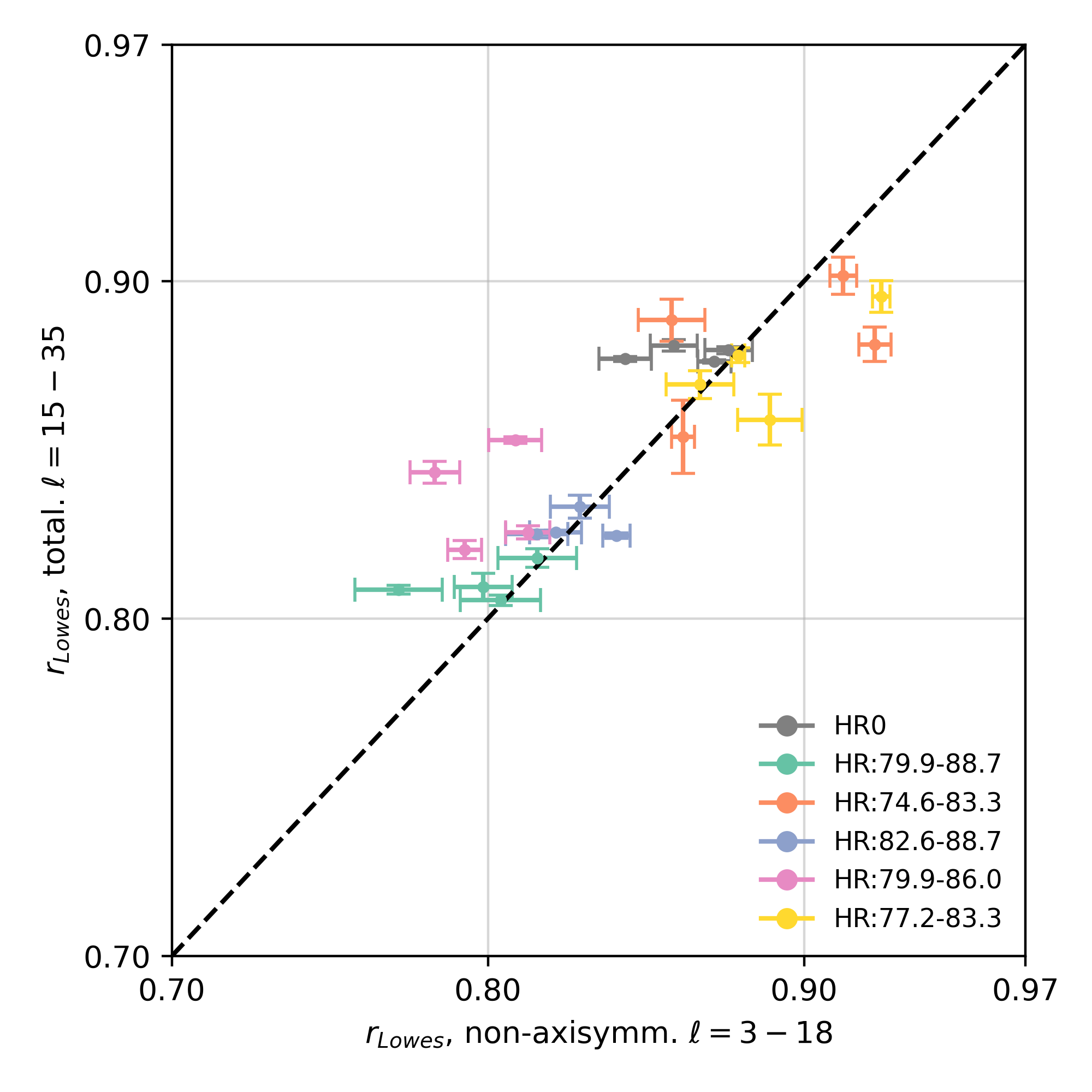}
\caption{Comparison of $r_{Lowes}$ obtained via fitting of the non-axisymmetric spectral components between $\ell=3-18$ (directly comparable to Jupiter measurements) and fitting of the total spectrum between $\ell=15-35$. \label{fig:CompLowesFits}}
\end{figure}

The Lowes radii inferred using the two fitting methods, $\ell=3-18$ (non-axisymmetric spectrum) and $\ell=15-35$ (total spectrum), in eq.~(\ref{eq:rlowes}), are given in Table~\ref{tab:Results}.


\begin{figure*}[t]
\centering\includegraphics[width=1.0\linewidth]{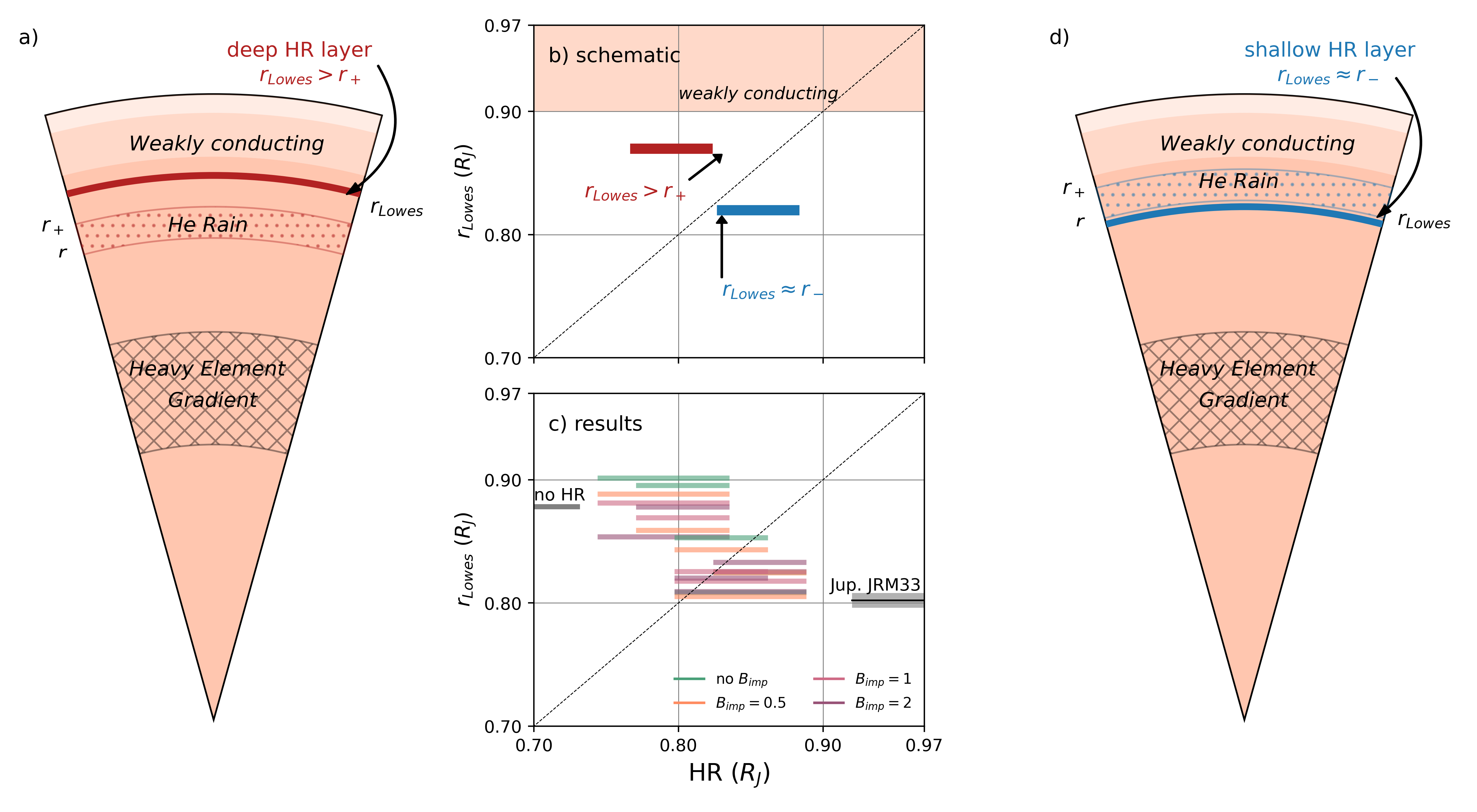}
\caption{a) and d) are schematic slices to put the results into the Jupiter context. They link with the two helium rain layer examples shown in schematic b) to aid in the interpretation of c), which shows inferred Lowes radius as a function of helium rain layer location, computed using $\ell=15-35$. Very similar $r_{Lowes}$ are obtained for models with no helium rain layer so an average is plotted (grey horizontal line). \label{fig:Lowes_Sims}}
\end{figure*}

The two fitting methods are compared in Fig.~\ref{fig:CompLowesFits}. In general, the error bars for the Lowes radii inferred using the $\ell=15-35$ fitting are smaller than for the lower degree fitting. Furthermore, comparing each set of models with the same helium rain layer location, which we would expect to have the same dynamo radius, there is less scatter for $r_{Lowes}$ obtained using the higher degree fitting. This suggests that the higher-degree components of the magnetic spectrum better represent the characteristic slope of the spectrum, while lower degrees could be more distorted by individual field characteristics. Unfortunately, no measured magnetic field spectrum at giant planets has been resolved to such high spatial resolution, making this fitting method unattainable for direct application at the planets. However, the inferred Lowes radii are in sufficient agreement that a fitting using the non-axisymmetric spectral components from $\ell=3-18$  still yields meaningful results.

In Fig.~\ref{fig:Lowes_Sims}c we illustrate how the inferred Lowes radii (using the fit to $\ell=15-35$ of the total spectra) are linked to the location of the helium rain layer for all of the models in this study. We use Fig.~\ref{fig:Lowes_Sims}b to aid in the interpretation of this figure, where we plot $r_{Lowes}$ as a function of helium rain layer location, with a horizontal line indicating the extent of the layer. We include a dashed diagonal line as this demonstrates how models with deep stably stratified layers (on the left side of the plot) tend to have shallower $r_{Lowes}$ as the inferred dynamo radius lies above the layer, while models with shallow stable layers have $r_{Lowes}$ at the lower boundary of the layer. In Fig.~\ref{fig:Lowes_Sims}b we also shade the region where electrical conductivity is negligible, thus acting as an upper limit to where the dynamo radius may be. Panels a) and d) place these results into the context of different Jovian structural models.
\begin{figure}[t]
\centering
\includegraphics[width=1.0\linewidth]{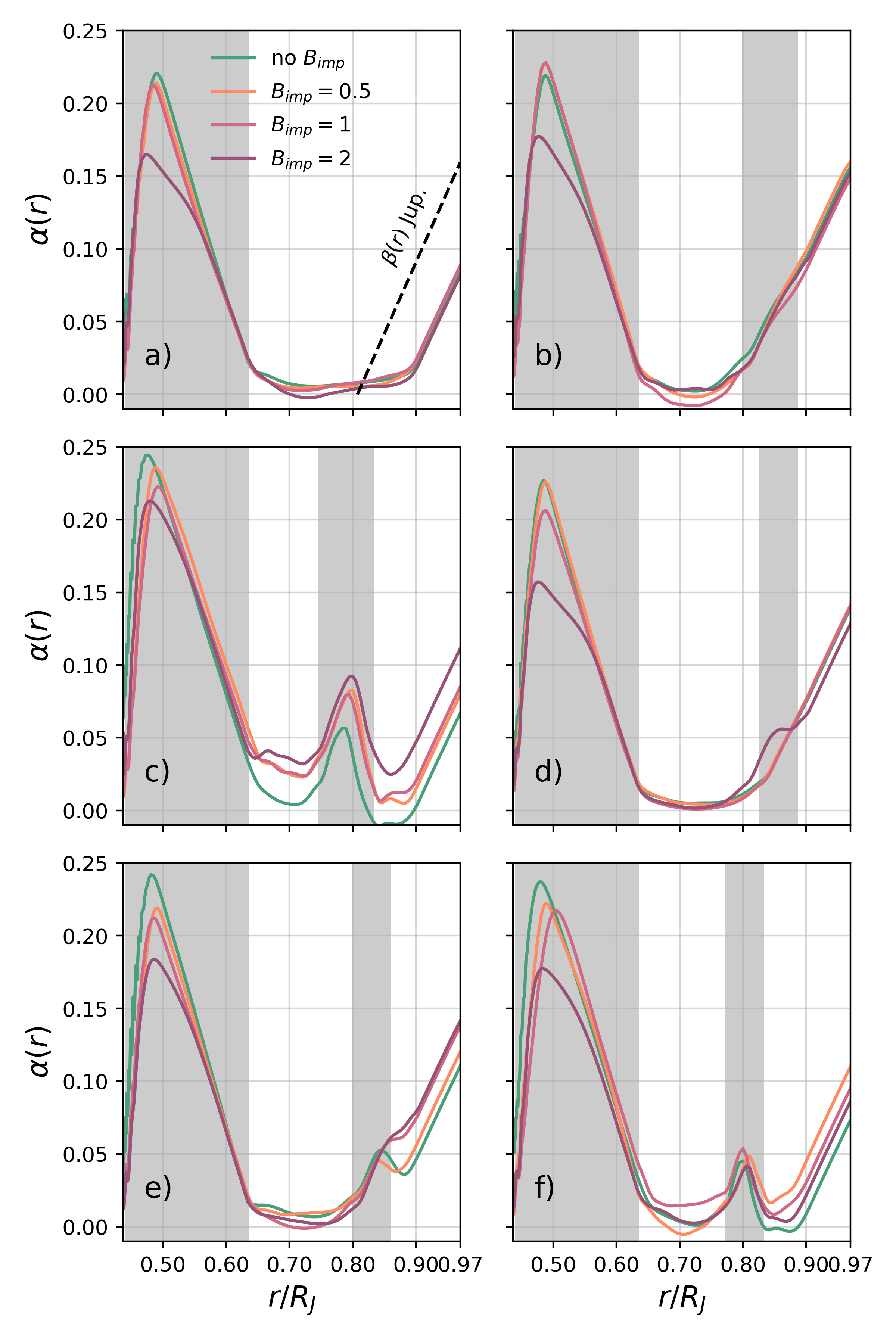}
\caption{Magnetic energy spectrum gradient, $\alpha(r)$, as a function of radius. We show the fitting for $\alpha(r)$ using degrees 15-35. In a) the black dashed line shows $\beta(r)$ for Jupiter's spectrum. Colour indicates the amplitude of the imposed axial dipole at the lower boundary while each panel contains models with a specific helium rain location, where grey shading shows where $d\tilde{s}/dr$ is positive, i.e. the subadiabatic region. \label{fig:alpha}}
\end{figure}

In order to investigate whether the actual ``top of the dynamo" may be considered to be below the helium rain layer, or at the depth at which electrical conductivity increases, we evaluate the slope of the magnetic energy spectrum as a function of radius (eq.~\ref{eq:alpha}, Fig.~\ref{fig:alpha}). For simplicity we show only $\alpha(r)$ profiles evaluated using $\ell=15-35$.

For models without a helium rain layer, a Lowes radius of just over 0.87 is inferred (an averaged $r_{Lowes}$ is represented by the grey horizontal line on the left in Fig.~\ref{fig:Lowes_Sims}c, as the results for all $B_{imp}$ were very similar). This is similar to previous, fully convective, models with a varying electrical conductivity that is also fitted to the Jupiter profile, such as \citet{Tsang_2020}, as the slope of the spectrum at the surface is mainly linked to the depth at which electrical conductivity increases. For these models, the Lowes spectrum fitting is adequate, as Fig.~\ref{fig:alpha}a shows that the slope of the spectrum $\alpha(r)$ does, indeed, reach very small values around this depth. However, as pointed out in \citet{Tsang_2020}, $r_{Lowes}$ represents a lower limit to the actual dynamo radius, as the spectrum does not become fully flat throughout most of the dynamo region. This led \citet{Tsang_2020} to define the dynamo radius of their models as the depth at which $\alpha(r)$ departs from $\beta(r)$. While in some models the gradient does fall to zero, the value of $\alpha$ in what may physically be considered the dynamo region is generally $<0.025$ (with the exception of HR74.6-83.3 models where the slope remains steeper).

The inferred Lowes radii are more difficult to interpret for models with a helium rain layer as illustrated by Fig.~\ref{fig:alpha}b-f. For models with deep helium rain layers, such as in Fig.~\ref{fig:alpha}c and f ($0.746-0.833R_J$ and $0.772-0.833R_J$, respectively) the curves for $\alpha(r)$ in the outer convective region are somewhat similar to those for the no helium rain layer cases. The inferred values of $r_{Lowes}(\ell=15-35)$ also tend to lie between $0.85-0.9R_J$, albeit with a little more variation between cases. The implication that there is field generation occurring above the helium rain layer is supported by the actual gradients of the spectra in this region as $\alpha(r>r_{HR})$ can be seen to fall to values similar to those in the deeper dynamo region.

For shallow helium rain layers, such as those with an upper boundary around $0.887R_J$ (shown in panels b and d) the top of the dynamo region lies at the lower boundary of the stably stratified layer. While the profiles of $\alpha(r)$ show that the slope of the spectrum does not follow that of a scalar potential exactly, it does not deviate dramatically and therefore the Lowes radius decently infers the depth at which the spectrum becomes almost flat. However, mid-depth layers are more complicated as Fig.~\ref{fig:alpha}e shows that with an upper boundary at $0.860R_J$ there is some field being generated above the helium rain layer, as the slope deviates strongly from that of a potential, in particular for models with weak imposed dipole fields, yet it is far from flattening out. The difference between the spectra for the four models with the HR layer between $79.9-86.0\%\,R_J$ may be attributed to the presence of a stronger magnetic field in models -B1 and -B2, responsible for inhibiting convection slightly.

\subsection{Secondary Dynamo Action}
\begin{figure}[h!]
\centering
\includegraphics[width=1.0\linewidth]{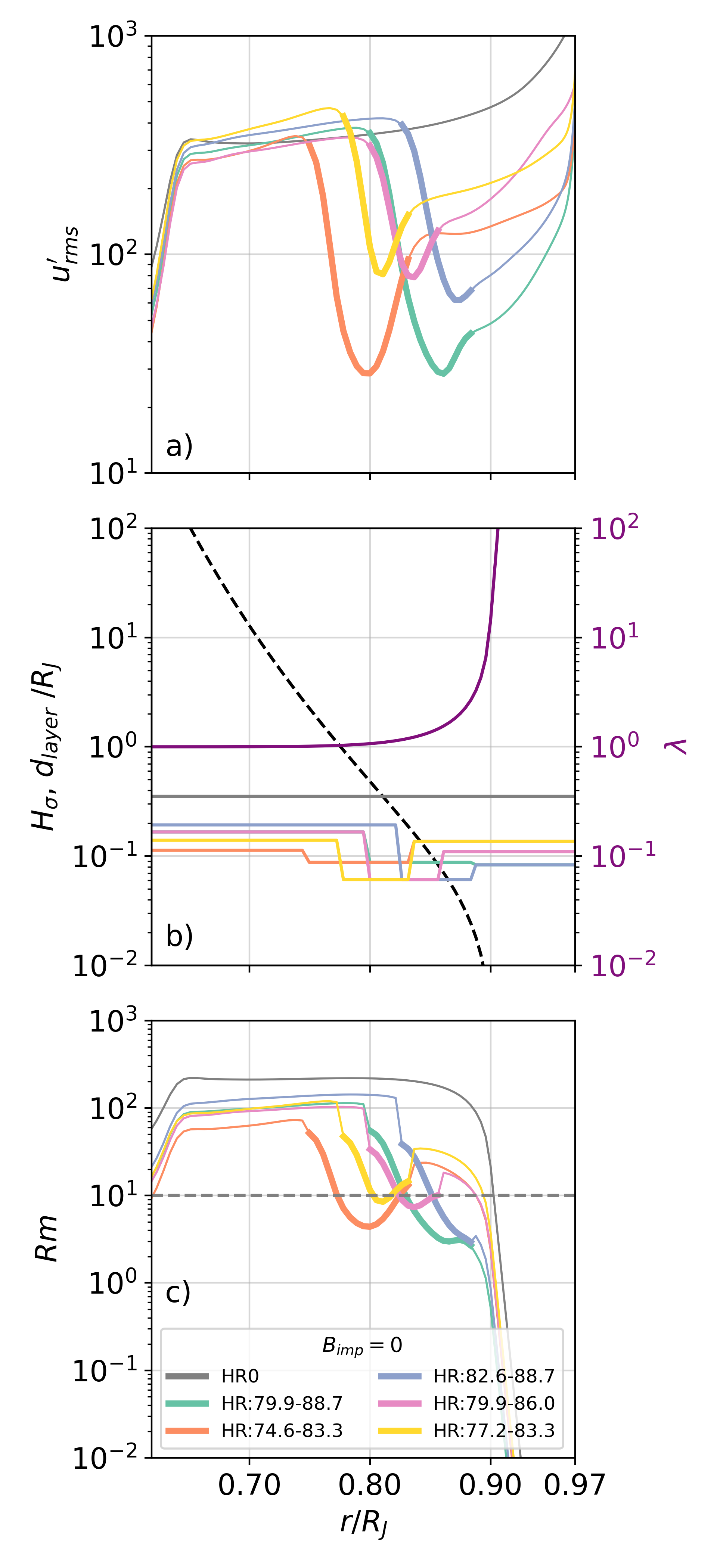}
\caption{a) Horizontally averaged rms non-axisymmetric velocity (eq.~\ref{eq:urms}), in units of $Re$, for the six models without an imposed dipole and varying helium rain layer location. b) Left axis: relevant length-scales i.e. electrical conductivity scale height $H_\sigma$ (black dashed line) and local layer thickness for each model; right axis: magnetic diffusivity. c) Local magnetic Reynolds number based on non-axisymmetric flow and eq.~\ref{eq:Rm}. Thicker lines are used in the regions where each model's helium rain layer lies. \label{fig:Rm}}
\end{figure}


To evaluate the strength of the local field generation leading to the differences observed in the spectral slopes, we consider the local $Rm$, as discussed in sec.~\ref{sec:Rm}, associated with the non-axisymmetric flows in our models:
\begin{equation}
    u^\prime_{rms}(r) = \frac{1}{4\pi}\oint\langle|u^\prime(r, \theta, \phi, t)|\rangle_t\sin\theta d\theta d\phi. \label{eq:urms}
\end{equation}
Here, $\langle...\rangle_t$ represents the time-average. In the shallower regions the electrical conductivity scale height $H_\sigma$ (Fig.~\ref{fig:Rm}b, dashed black line), as defined in sec.~\ref{sec:Rm}, is typically the relevant length scale as it is much smaller than the layer thickness. However, as the convective outer envelope is interrupted by the presence of a helium rain layer in our models, this is not the case below $0.9\,R_J$. Therefore, when evaluating $Rm$ in our models, we use the local layer thickness $d_{layer}$ as the relevant length-scale. As shown in Fig.~\ref{fig:Rm}b, this is specific for each model and based on the distance between the upper and lower boundaries of each layer:
\begin{equation}\label{eq:dlayer}
    d_{layer}(r)=\begin{cases}
        r_{HR} - r_{HEG}, & \text{if $r_{HEG}<r<r_{HR}$},\\
        d_{HR}, & \text{if $r_{HR}<r<r_{HR}+d_{HR}$},\\
        r_o-(r_{HR}+d_{HR}), & \text{if $r_{HR}+d_{HR}<r<r_o$}.
  \end{cases}
\end{equation}
The magnetic Reynolds number radial profiles shown in Fig.~\ref{fig:Rm}c are thus computed as:
\begin{equation}
    Rm= \frac{u^\prime_{rms}d_{layer}}{\lambda}.\label{eq:Rm}
\end{equation}
For the shallow helium rain layer models, with an upper boundary of $88.7\%\,R_J$, $Rm$ does not exceed 10 above the HR layer. This explains why there is barely any deviation from a potential from the bottom of the HR layer to the model surface.

The intermediate model with an upper boundary of $86.0\%\,R_J$ surpasses $Rm=10$ above the HR layer, reaching a maximum value of slightly less than 20. This indicates that there is some magnetic field being induced by the convective motions in the region between $0.860<r/R_J<0.9$ and leads to the complex spectral slope profile observed in Fig.~\ref{fig:alpha}e.

In the deep HR layer models $Rm\approx30$ in the region $0.820<r/R_J<0.9$, providing clear evidence of a secondary dynamo region as suggested by the spectral slopes. It is possible that this secondary dynamo only operates in the presence of the strong magnetic field generated below, as pointed out by \citet{Gastine_2014} and \citet{Cao_2017}. However, there is no easy way to show this explicitly.

\section{Discussion and Conclusion}

\subsection{Interpretation at Jupiter}

Applying the Lowes method to models with a Jovian electrical conductivity profile and with a shallow stably stratified layer, we find the following:
\begin{itemize}
    \item For deeper helium rain layers a secondary dynamo is active above the stable layer. The upper boundary of this secondary dynamo is then linked directly with the electrical conductivity profile (see Fig.~\ref{fig:Lowes_Sims}a) and corresponds well with the inferred $r_{Lowes}$.
    \item For shallower helium rain layers, there are no significant electrical currents above the helium rain layer. The field follows a potential through to the bottom of the stable layer, leading to an inferred $r_{Lowes}$ at the base of the helium rain layer, which is also the top of the electrically conducting, convective region (see Fig.~\ref{fig:Lowes_Sims}d).
\end{itemize}
The fitting using the existing Juno measurements still suffers from the limitation of the spatial resolution, as our results show an increased reliability in the method when including degrees beyond $\ell=18$. However, Fig.~\ref{fig:CompLowesFits} suggests that using the non-axisymmetric components yields good results for a spectrum resolved up to $\ell=18$. For our numerical dynamo models, excluding the axisymmetric part was vital, as seen in the spectra in Fig.~\ref{fig:B0_SpecDecomp}. However, for JRM33 it does not make much difference as we obtain $r_{Lowes}= 0.8069\pm0.0059$ (non-axisymm., $\ell=3-18$) and $r_{Lowes}= 0.8074 \pm 0.0052$ (total spectrum). This is due to the axisymmetry of the Jovian spectrum being lower than many of our models. This is also an indication that the difference in the results of \citet{Connerney_2022} and those of \citet{Sharan_2022} are not due to the exclusion of the zonal components of the spectrum when applying the Lowes analysis to their models. Rather, they are due to differences in their methods of deriving the respective Jupiter magnetic field model from the Juno measurements. We did not test including degrees 1 and 2 in our fitting, as \citet{Sharan_2022} do. Fig~\ref{fig:B0_SpecDecomp} indicates that including the non-zonal part of the $\ell=2$ component of the spectra would not alter the fitting significantly. However, including the equatorial dipole, $m=1, \ell=1$, component could bias the fitting as Fig~\ref{fig:B0_SpecDecomp} shows that the equatorial dipole is significantly weaker than the other low degrees non-zonal components for all models plotted.

If the fitting of the slope between $\ell=3-18$ is taken as sufficiently representative of the gradient of the whole spectrum, the value of $r_{Lowes}=0.8069R_J$ (or $0.830\pm0.022R_J$ from \citet{Sharan_2022}) indicates the presence of a stably stratified layer with an upper boundary not much deeper than $\sim0.9R_J$ and extending to around $0.81-0.83\,R_J$. This is implied by the set of models with HR79.9-88.7, all with $r_{Lowes}\approx0.8R_J$, and the set with HR82.6-88.7, with $r_{Lowes}\approx0.82R_J$. The set of models with HR82.6-88.7, i.e. the same upper boundary but thinner, have slightly shallower $r_{Lowes}$, as magnetic field generation sets in right below the layer. Therefore, our results strongly suggest an interior structure similar to that shown in Fig.~\ref{fig:Lowes_Sims}d. However, the physical origin of this extended “shallow” stable layer and its connection to helium rain remain to be elucidated as the upper boundary of this layer appears to be deeper than the expected initiation depth of helium rain inside Jupiter.


Deeper HR layer models host a secondary dynamo region above the stable layer and therefore feature shallower $r_{Lowes}$, which are incompatible with that inferred from observed Jupiter spectrum. Therefore, our Lowes radius analysis results are not compatible with the extremely thin helium rain layer scenario favored by \citet{Markham_2024}. If such a thin stably stratified layer exists between $80-85\%~R_J$, we would expect dynamo action above the layer and a shallower value of $r_{Lowes}$.


The 3D numerical dynamo simulations we have carried out are quite rich in physics and we intend to make an additional study exploring the fluid dynamics and magnetic field morphologies of these models, their dependence on helium rain layer location, and their Jupiter-likeness.

\subsection{Application at Saturn}

As shown in Fig.~\ref{fig:B0_SpecDecomp}, the axisymmetric part of planetary magnetic spectra do not typically follow simple power-law distributions. Furthermore, the low-degree part of the axisymmetric spectra tend to exhibit strong oscillations which is both the case in our numerical dynamo models and at Saturn \citep{Dougherty_2018, Cao_2020}. Thus, the Lowes method cannot be easily applied to the extremely axisymmetric magnetic field of Saturn \citep{Dougherty_2018,Cao_2020}. One approach for a future Saturn study could be to use Saturn's quadrupole family spectrum, i.e. the $m+\ell=$even components, as considered by \citet{Langlais_2014}, with the caveat that only the axisymmetric part of the quadrupole family spectrum are accessible in the currently available observations at Saturn. 

\subsection{Application at the Ice Giants}

Our study represents an exciting prospect for the ice giants. If a future mission is able to resolve the magnetic fields of Uranus or Neptune to a high spherical harmonic degree, the Lowes method should provide a robust method of inferring their dynamo radius. This is indicated by the results for our models with a low dipolarity (e.g., models HR82.6-88.7-B0 and HR79.9-86.0-B0, see Table~\ref{tab:Results}), as the ice giants have. The Lowes method works equally well for these non-dipolar models.

\citet{Langlais_2014} made an attempt at applying it to the ice giants with the existing field model derived from data with limited spatial coverage. Our results show that applying the Lowes method to a sufficiently well-resolved spectrum of the ice giants' magnetic fields will also help distinguish between structure models which are fully convective in the outer regions, e.g. \citet{Militzer_2024b}, and those which include a $\text{H}_\text{2}–\text{H}_\text{2}\text{O}$ immiscibility layer in the region where electrical conductivity is increasing, e.g. \citet{Amoros_2024, Gupta_2025}, analogous to the helium rain layer inside Jupiter and Saturn.

\subsection{The Lowes method}

Despite being based on two broad assumptions, we show that the Lowes method works remarkably well in inferring the correct dynamo radius based on surface magnetic spectra. We find that the high degree spectral components are well characterized by the dynamo region's depth while low degree components, in particular the dipole, quadrupole and octupole, can be more affected by individual magnetic field properties. However, these are mostly imprinted on the axisymmetric part of the spectrum, in particular in the presence of a shallow stably stratified region, and applying the Lowes method to the non-axisymmetric spectrum can mitigate this effect.

While the magnetic spectrum does not become completely flat in the dynamo region (see Figs.~\ref{fig:alpha} and \ref{fig:EM_RLowes}), as also found by \citet{Tsang_2020}, it does have a very shallow gradient. The flatness of the spectrum, i.e. the validity of the White Source Hypothesis, needs to be further explored in future studies by varying the MHD parameters, such as the Rayleigh number. While a higher Rayleigh number will increase the turbulence of the system, it is not clear how this will affect the spectrum. Furthermore, exploration of different MHD parameter combinations are also likely to yield models which have a more Jupiter-like magnetic field morphology, including the level of axisymmetry.

When reviewing the Lowes approach of inferring the dynamo radius of a planet, we also consider that the general form of the poloidal magnetic spectrum in a planet's dynamo region may also be written as:
\begin{equation}
    R_\ell(r)=\text{const.}\,\ell^{-\gamma},
\end{equation}
where we note that $k$ is commonly used, rather than $\ell$, when considering the wavenumbers in MHD turbulence literature. The most well-known example is Kolmogorov's scaling for the kinetic energy density spectrum in the inertial range of a turbulent flow, where $\gamma=5/3$. Thus, perhaps we should not expect a completely flat spectrum at the dynamo surface, corresponding to $\gamma=0$. This has been explored in the context of Earth in Appendix B of \citet{Voorhies_2004}, as well as \citet{Roberts_2003, Buffett_2007}. It is then not surprising that the slope evaluated in this work does not reach zero in the dynamo region (Fig.~\ref{fig:alpha}). We propose a future study in the gas planet context, that explores the possibility of a spectrum with $\gamma\neq0$ and its impact on the inferred value of the dynamo radius.

\appendix
\section{Time-averaged Lowes-Mauersberger Spectra}\label{sec:AppA}

\begin{figure}[t!]
\centering
\includegraphics[width=\linewidth]{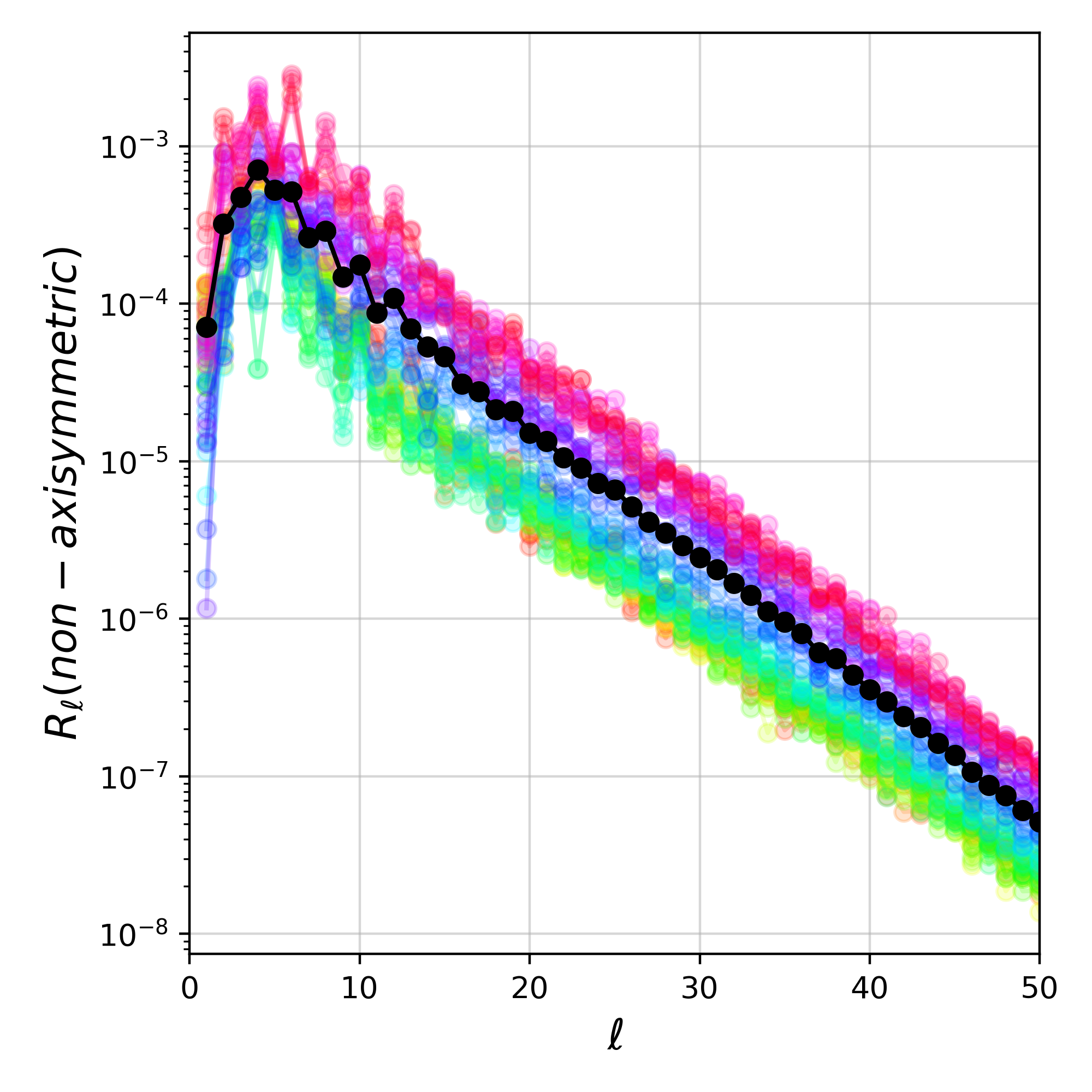}
\caption{Lowes-Mauersberger spectra for model HR0-B0. Colored lines are instantaneous spectra while the black line shows the averaged spectrum, used for the Lowes analysis in this study. \label{fig:SpecSnaps}}
\end{figure}

Our Lowes radius analyses were performed on time averaged $R_\ell$ spectra. Each spectrum was computed as the average of individual instantaneous $R_\ell$ spectra. (Thus, our spectra are not calculated based on an time-averaged magnetic field.) This is shown in Fig.~\ref{fig:SpecSnaps} for model HR0-B0, where the colored spectra are snapshots of the non-axisymmetric component of the Lowes-Mauersberger spectrum and the black line is the time averaged spectrum on which the Lowes method was applied. While there is some natural fluctuation of the amplitude of the field over the time interval \citep[which is to be expected, see][for example]{Amit_2010}, the slopes, $d R_\ell / d \ell$, of the instantaneous spectra and their time average are all closely consistent.

\section{Magnetic Field Spectra at the inferred Lowes Radii}\label{sec:AppB}

\begin{figure}[h]
\centering
\includegraphics[width=\linewidth]{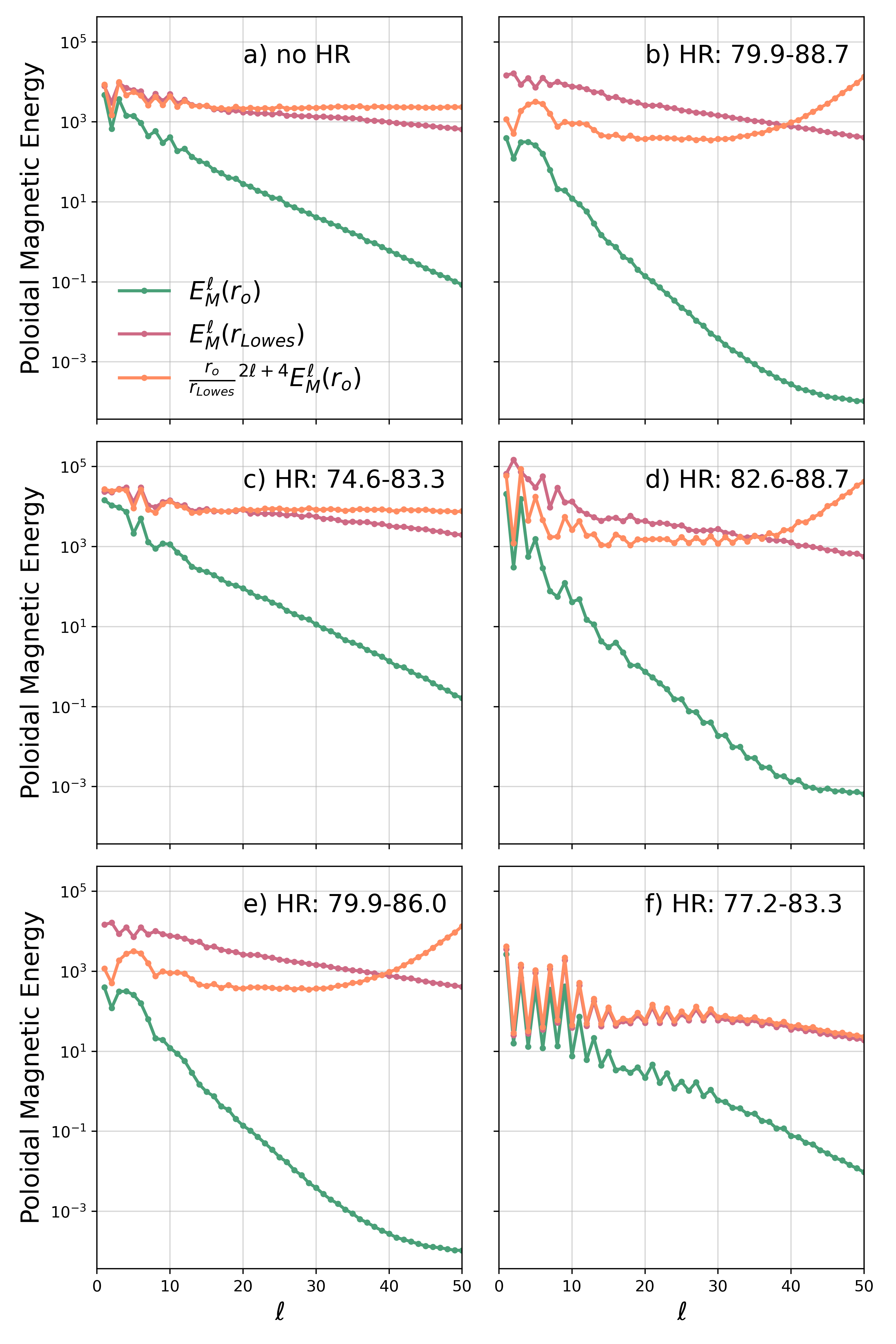}
\caption{Time-averaged poloidal magnetic energy spectra for 6 models with no imposed axial dipole field and varying helium rain layer locations. Green lines correspond to the spectrum spectra at the model surface ($r_o=0.97R_J$), pink lines are the spectra at each model's Lowes radius, and orange is the surface spectrum extrapolated down to the inferred Lowes radius. \label{fig:EM_RLowes}}
\end{figure}

Fig.~\ref{fig:EM_RLowes} shows the poloidal magnetic energy spectra for the suite of models with no imposed axial dipole field. The surface spectra are shown in green. These spectra at $r_o$ are extrapolated as a potential field down to the respective model's Lowes radius in orange. Additionally, the actual poloidal magnetic field spectra at the inferred Lowes radius are shown in pink for comparison.

For the model without a stable layer (Fig.~\ref{fig:EM_RLowes}a) and those with deep stable layers (Fig.~\ref{fig:EM_RLowes}c, and f), the spectra downward-continued from the surface and the actual spectra at the Lowes radius are very similar, i.e. $E_M^\ell(r_{Lowes})\approx\frac{r_o}{r_{Lowes}}^{2\ell+4}E_M^\ell(r_o)$.

However, in models where the inferred Lowes radius is below the stable layer (Fig.~\ref{fig:EM_RLowes}b and d) or within the layer (Fig.~\ref{fig:EM_RLowes}e) the spectrum at $r_{Lowes}$ deviates strongly from that of the extrapolated surface spectrum. This indicates that while the slope of the spectrum is sufficiently similar to that of a potential when passing through the stable layer for the Lowes method to work, the characteristics of the spectrum are altered significantly.

\section*{Acknowledgments}
PNW and HC acknowledge support from the NASA Juno project and the NASA CDAP grant No. 80NSSC23K0511. JMA thanks the NSF Geophysics Program for support via award EAR \#2143939. We thank Hagay Amit for his thoughtful review and suggestions which significantly improved the manuscript.

\bibliographystyle{aasjournal}


\end{document}